\documentclass[epj]{svjour}

\usepackage{bm}   
\usepackage{color}
\usepackage{amssymb}
\usepackage{amsmath}
\usepackage{balance}
\usepackage{subfigure}
\usepackage{graphicx}
\usepackage{amsfonts}
\usepackage[colorlinks=true, linkcolor=blue, citecolor=blue, urlcolor=magenta]{hyperref}
\usepackage{algorithmic}
\usepackage{algorithm}
\usepackage{multirow}
\usepackage{array}
\usepackage{booktabs}
\usepackage{xcolor}
\usepackage{siunitx}

\title{Multivariate time series prediction using clustered echo state network}
\author{S. Hariharan, R. Suresh{$^*$}, V. K. Chandrasekar}

\institute{$^{1}$Department of Physics, Centre for Nonlinear Science and Engineering, School of Electrical and Electronics Engineering,\\ SASTRA Deemed University, Thanjavur 613 401, India. \\\\$^*$Corresponding author email: suresh@eee.sastra.edu}

\date{Received: -- / Revised version: --}

\abstract{
	Many natural and physical processes can be understood by analyzing multiple system variables evolving, forming a multivariate time series. Predicting such time series is challenging due to the inherent noise and interdependencies among variables. Echo state networks (ESNs), a class of Reservoir Computing (RC) models, offer an efficient alternative to conventional recurrent neural networks by training only the output weights while keeping the reservoir dynamics fixed, reducing computational complexity. We propose a clustered ESNs (CESNs) that enhances the ability to model and predict multivariate time series by organizing the reservoir nodes into clusters, each corresponding to a distinct input variable. Input signals are directly mapped to their associated clusters, and intra-cluster connections remain dense while inter-cluster connections are sparse, mimicking the modular architecture of biological neural networks. This architecture improves information processing by limiting cross-variable interference and enhances computational efficiency through independent cluster-wise training via ridge regression. We further explore different reservoir topologies, including ring, Erd\H{o}s--R\'enyi (ER), and scale-free (SF) networks, to evaluate their impact predictive performance. Our algorithm works well across diverse real-world datasets such as the stock market, solar wind, and chaotic R\"ossler system, demonstrating that CESNs consistently outperform conventional ESNs in terms of predictive accuracy and robustness to noise, particularly when using ER and SF topologies. These findings highlight the adaptability of CESNs for complex,  multivariate time series forecasting.
	\PACS{
		{PACS-key}{ }
	} 
}
\titlerunning{Noise-induced extreme events}
\authorrunning{S. Hariharan et al.}
\begin{document}
	
	\maketitle
	\section{Introduction}
\label{sec1}
Echo state network (ESN), a subset of reservoir computing (RC), provides an efficient computational framework for processing sequential data, particularly in time series prediction and pattern recognition \cite{jaeger2001echo,jaeger2002adaptive}. ESNs comprise an input layer, a dynamic reservoir (consisting of nodes, spiking neurons, or other dynamical systems), and an output layer. The reservoir maps input data into a high-dimensional space, enabling efficient learning while significantly reducing computational complexity. Unlike traditional recurrent neural networks (RNNs), ESNs avoid full backpropagation and train only the readout layer \cite{sun2020review}. The stability of ESNs relies on the echo state property, which is governed by specific hyperparameters \cite{yan2024emerging}.

Since the reservoir in a standard ESN is a randomly connected network, extensive research has explored different connectivity structures to enhance its performance. Given that artificial neural networks (ANNs) are designed to mimic brain functionality, studies have investigated the effects of clustering nodes within the reservoir using various algorithms. Several approaches have been proposed, including scale-free, highly clustered ESNs \cite{deng2007collective}, clustering of scale-free and small-world networks using centroid, Ward, and K-Medoids methods \cite{najibi2015scesn}, a priori data-driven multi-clustered ESNs \cite{li2015priori}, and clustered ESNs based on Erd\H{o}s--R\'enyi (ER) and Barab\'asi-Albert (BA) networks for signal denoising and filtering \cite{junior2020clustered}. In addition, ESNs have been structured to resemble cortex-like networks, mimicking mammalian brain function \cite{song2010effects}. More recently, various reservoir topologies, including ring, lattice, and torus networks, have been explored to enhance network performance \cite{mcallister2024topological}. Clustered ESNs have also been applied to medical image classification tasks \cite{arroyo2023modified}. However, most existing studies have primarily focused on univariate time series prediction tasks, such as chaotic time series forecasting, traffic network analysis \cite{yu2011clustered}, and financial time series modeling \cite{xue2016application}, with little emphasis on multivariate time series prediction. The major existing approaches are systematically compared in terms of their clustering strategies, inter-cluster connection modes, and training methodologies to clearly emphasize the novelty of the proposed algorithm, as summarized in Table \ref{tab:clustered_esn_comparison}.

\begin{table}[htbp]
	\centering
	\renewcommand{\arraystretch}{1.2}
	\setlength{\tabcolsep}{5pt}
	\begin{tabular}{p{3.5cm} p{4.5cm} p{5.5cm} p{2.5cm}}
		\hline
		\textbf{Method} & \textbf{Clustering Basis} & \textbf{Inter-Cluster Connection Mode} & \textbf{Training} \\
		\hline
		Deng \& Zhang (2007) -- \textit{Small-world scale-free RNN} \cite{deng2007collective}
		& Cluster formation via network evolution (Topology growth rules) 
		& Hierarchical sparse links -- fixed at construction 
		& Global readout \\
		
		Najibi \& Rostami (2015) -- \textit{SCESN / SPESN / SWESN} \cite{najibi2015scesn}
		& Data-driven clustering based on K-Means, PAM \& Ward algorithms 
		& Backbone + local structure -- small-world (SW) style, fixed at construction 
		& Global readout \\
		
		Li et al. (2015) -- \textit{Data-driven multi-cluster ESN} \cite{li2015priori}
		& A priori data-based clustering -- variable and feature corresponding 
		& Inter-cluster links based on data priors, \& fixed at generation 
		& Global readout\\
		
		Oliveira Jr. et al. (2020) -- \textit{Clustered ER / BA ESN} \cite{junior2020clustered}
		& Clustered graph models (ER, BA) -- topology-based 
		& Intra-cluster dense + sparse inter-cluster -- fixed graph structure 
		& Global readout\\
		
		McAllister et al. (2024) -- \textit{Topological \& simplicial RC} \cite{mcallister2024topological}
		& Topological/simplicial structure -- topology-based 
		& Static topological patterns -- analyzed, not adaptive 
		& Global readout\\
		
		Proposed Clustered ESN for multivariate time series 
		& No. of clusters = No. of variables to be predicted; Inputs mapped specific to the clusters;
		& Sparse inter-cluster connection and dense intra-cluster connections 
		& Separate readout for each cluster \\
		\hline
	\end{tabular}
	
	\caption{Comparison of clustering principles, inter-cluster connection schemes, and training strategies adopted in various clustered or structured reservoir computing frameworks.}
	\label{tab:clustered_esn_comparison}
\end{table}

A multivariate time series consists of multiple interdependent variables representing different aspects of a system, where each variable may influence others directly or indirectly. Predicting such time series is a challenging task due to the complex interactions between variables, and several machine learning models have been developed to address this problem. Various modifications to ESNs have been introduced for multivariate time series prediction, including adaptive elastic ESNs with regularization techniques \cite{xu2016adaptive}, robust variational ESNs \cite{shen2018novel}, broad ESNs \cite{yao2019broad}, hybrid regularized ESNs \cite{xu2018hybrid}, and Laplacian ESNs \cite{han2017laplacian}. More recently, biogeography-based optimization has been proposed for optimizing the hyperparameters \cite{na2020modified}, and hierarchical ESNs have been applied to multidimensional chaotic time series prediction \cite{na2022hierarchical}. Additionally, modified binary salp swarm algorithm-based ESNs \cite{ren2022multivariate} have been introduced to further enhance performance. While these approaches improve predictive accuracy, they introduce significant complexity, particularly due to the need for extensive hyperparameter tuning. Some methods, such as broad ESNs \cite{yao2019broad} and deep ESNs \cite{malik2016multilayered}, employ multiple reservoirs, making the optimization process more challenging.

To overcome these limitations, we introduce a clustered ESN (CESN) framework that efficiently captures multivariate temporal dependencies within a single clustered reservoir. Unlike conventional ESNs, which randomly distribute connections within the reservoir, our approach clusters reservoir nodes based on the number of variables in the multivariate time series. Each input variable is directly mapped to a designated cluster, ensuring that every cluster is responsible for processing a single component of the multivariate time series. This structured organization enhances information processing by minimizing cross-variable interference while retaining the inherent advantages of the reservoir computing paradigm. The intra-cluster weights are initialized from a defined distribution, whereas the inter-cluster links are intentionally kept sparse to emulate modular, brain-like network connectivity \cite{meunier2010modular,hilgetag2000anatomical}. Since inputs are directly mapped to their corresponding clusters, training is performed independently within each cluster, and the readout layer learns the output weights through ridge regression using the respective reservoir states. To evaluate the impact of reservoir topology, we explore different clustered network configurations, including ring, ER, and SF networks. The connection densities of these reservoirs are quantitatively evaluated to assess their impact on predictive performance across various configurations.

To validate the effectiveness of the proposed CESN approach, we employ it on two real-world multivariate time-series datasets: the Indian National Stock Exchange (NSE) stock-market data \cite{nse_all_reports}, and the solar-wind data acquired from NASA’s DISCOVR satellite \cite{noaa_solar_wind}. These datasets are inherently noisy, making their prediction particularly challenging. While ESNs have been widely used for real-world time series forecasting \cite{sun2020review}, handling such noisy multivariate data remains an open research problem. To further demonstrate the robustness of the proposed algorithm, a benchmark evaluation is conducted on the chaotic Rössler system, which is well known for its high sensitivity to initial conditions and nonlinear dynamics. Through these comprehensive experiments, this study bridges the existing research gap by demonstrating that a clustered-reservoir architecture can substantially enhance predictive accuracy for complex multivariate time series while preserving computational efficiency.

The remainder of this paper is organized as follows. Section \ref{sec2} provides an overview of the traditional ESN architecture and its training methodology. In Section \ref{sec3}, we introduce the proposed clustered reservoir framework and present a detailed quantitative evaluation of its performance. Section \ref{sec4} offers a comprehensive assessment of the predictive capabilities of the clustered ESN across two real-world multivariate time series datasets. In Section \ref{sec5}, we apply the proposed method to the chaotic R\"ossler dynamical system to demonstrate its robustness and generalizability. Finally, Section \ref{sec6} summarizes the key findings and discusses potential directions for future research.

\section{Steps involved during the training process in conventional ESN}
\label{sec2}
\begin{figure}[h!]
	\centering
	\includegraphics[width=0.6\linewidth]{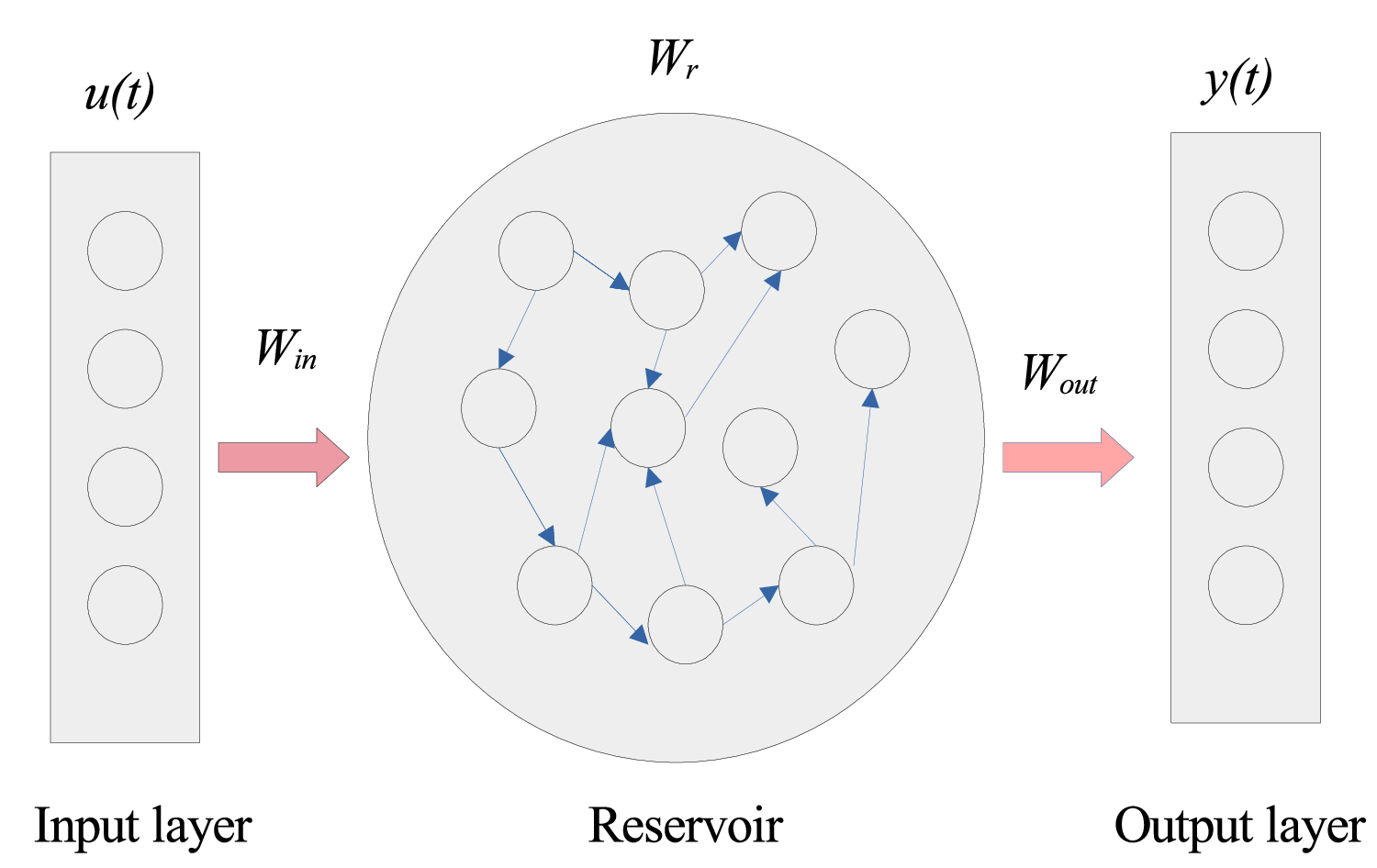}
	\caption{Basic architecture of the Echo state network}%
	\label{esn_arc}
\end{figure}

ESNs consist of three main components: an input layer, a dynamic reservoir, and an output layer. The input layer processes an input vector $\mathbf{u}(t)$, where the number of input variables to be predicted determines the dimension of this vector ($N_u$) at time $t$. This input is mapped into the reservoir, which comprises a set of $N_r$ nodes forming the computational core of the ESNs. Finally, the output layer generates the predicted output vector $\mathbf{y}(t)$, with $N_y$ dimensions corresponding to the number of predicted features. The fundamental architecture of an ESN is illustrated in Fig. \ref{esn_arc}.

The ESN operates through a set of functional weight matrices that govern the interactions between its layers. The input weight matrix, $\mathbf{W}_{\text{in}} \in \mathbb{R}^{N_r \times N_u}$, connects the input vector $\mathbf{u}(t)$ to the reservoir. The reservoir weight matrix, $\mathbf{W_r} \in \mathbb{R}^{N_r \times N_r}$, defines the internal connections between reservoir nodes, with its values typically drawn from a random distribution in the range [-1,1]. Lastly, the output weight matrix, $\mathbf{W}_{\text{out}} \in \mathbb{R}^{N_y \times N_r}$, maps the reservoir states to the output.

Before training, the input sequence $\mathbf{u}(t) \in \mathbb{R}^{N_u}$ is preprocessed, which may involve normalization and scaling to align with the dynamic range of the reservoir. The reservoir weight matrix $\mathbf{W_r}$ is initialized as a sparse, randomly connected network. It is then scaled to ensure that its spectral radius ($\rho$) is less than 1 ($\rho < 1$), thereby maintaining the echo state property \cite{jaeger2002adaptive}. The input weight matrix $\mathbf{W_{in}}$ is also initialized to control the influence of the input signal on the reservoir dynamics.

At each time step $t$, the reservoir state $\mathbf{x}(t) \in \mathbb{R}^{N_r}$ is updated according to:

\begin{eqnarray}
	\label{eq: eq1}
	\mathbf{x}(t) = (1 - \alpha)\mathbf{x}(t-1) + \alpha \tanh\left(\mathbf{W_{in}} \mathbf{u}(t) + \mathbf{W_r} \mathbf{x}(t-1) \right),
\end{eqnarray}

where $\alpha \in [0, 1]$ is the leak rate, which controls the temporal memory of the reservoir. The activation function $\tanh$ is applied element-wise to introduce nonlinearity.

The final network output $\mathbf{y}(t) \in \mathbb{R}^{N_y}$ is computed as a linear combination of the input and reservoir states:
\[
\mathbf{y}(t) = \mathbf{W_{out}}  \mathbf{x}(t) 
\]
where $\mathbf{W_{out}} \in \mathbb{R}^{N_y \times (N_u + N_r)}$ represents the trainable output weight matrix. These output weights are learned by solving a ridge regression problem \cite{marquardt1975ridge}, formulated as:

\begin{equation}
	\min_{\mathbf{W_{out}}} \|\mathbf{Y} - \mathbf{W_{out}} \mathbf{X}\|^2 + \lambda \|\mathbf{W_{out}}\|^2,
\end{equation}

where $\mathbf{Y}$ is the target output matrix, $\mathbf{X}$ is the concatenated matrix of inputs and reservoir states, and $\lambda$ is the regularization parameter, which prevents overfitting. The ridge regression approach ensures efficient training, as only the output weights $\mathbf{W_{out}}$ are adjusted while the reservoir weights remain fixed, making the approach computationally efficient. However, the conventional ESN structure can be further improved by incorporating a clustered reservoir architecture, where reservoir nodes are grouped into clusters based on the number of input variables. This modification, which enhances the model's capacity for multivariate time series prediction, is discussed in detail in the next section.

\section{Clustered reservoir}
\label{sec3}

\begin{figure}[h!]
	\centering
	\includegraphics[width=0.6\linewidth]{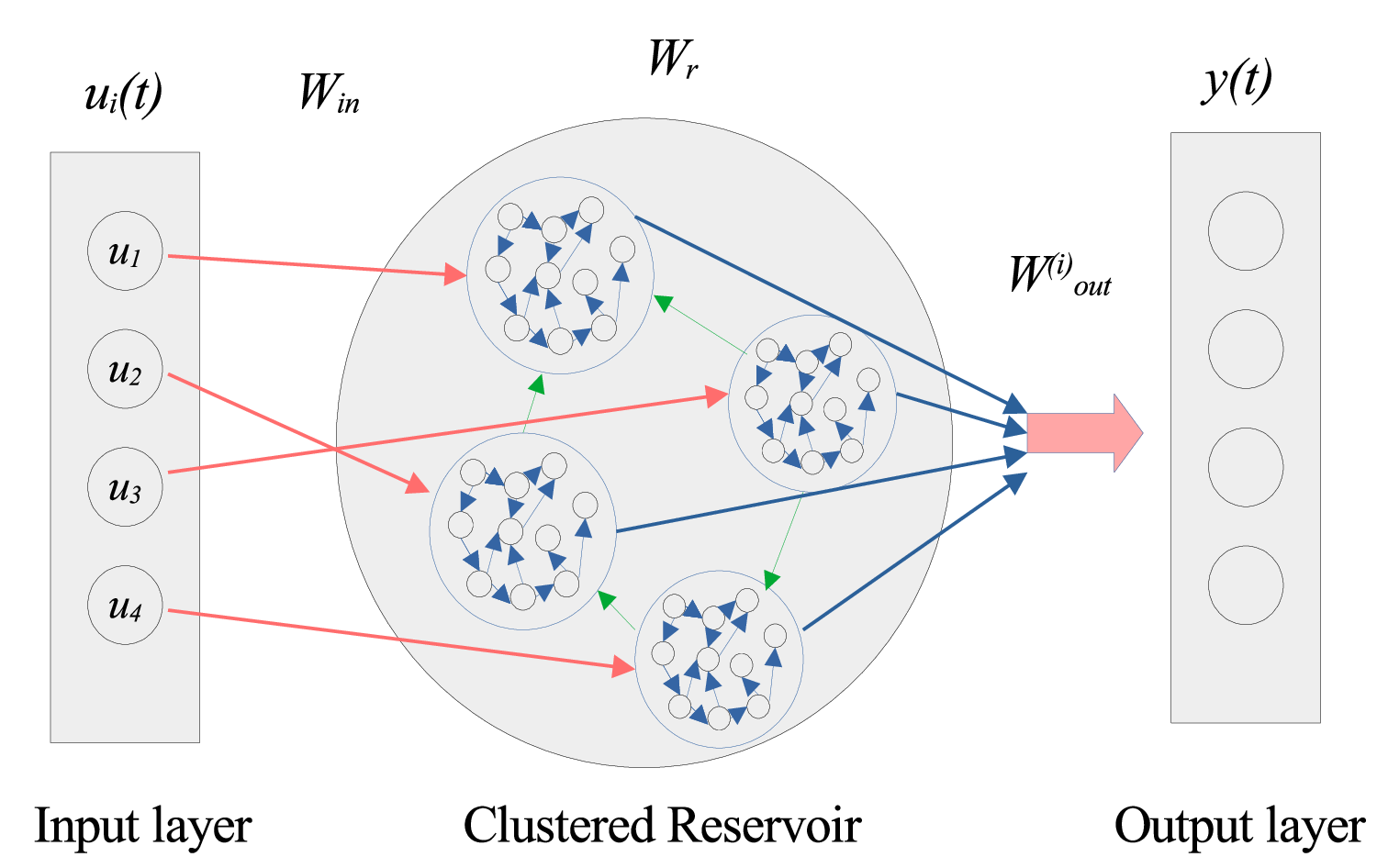}
	\caption{Schematic representation of the CESN architecture, where inputs from a multivariate time series are directed to specific reservoir clusters. Each cluster processes different aspects of the input dynamics, enhancing feature extraction and representation. }%
	\label{cluster}
\end{figure}
In this section,  we introduce the concept of clustering and demonstrate how inputs can be mapped to specific clusters. The clustered echo state network (CESN) is designed with the following structure: The reservoir consists of $N_{\text{r}}$ neurons, which are divided into $m$ clusters. Each cluster contains $N_{\text{c}}$ neurons, such that $N_{\text{r}} = m \cdot N_{\text{c}}$. Sparse inter-cluster connections enable limited communication between clusters, a characteristic observed in the modular architecture of the mammalian brain \cite{meunier2010modular}. In the CESN framework, inputs are selectively directed to specific clusters through the input weight matrix $\mathbf{W}_{\text{in}}$, which maps input signals to the reservoir states. The mapping process is as follows:

Each variable in the multivariate time series is associated with a specific cluster, assuming that the number of predicted variables equals the number of clusters ($m$). Consequently, each cluster is responsible for processing its designated input signal. The overall architecture of the clustered reservoir is illustrated in Fig. \ref{cluster}.

The input weight matrix is initialized as $\mathbf{W}_{\text{in}} \in \mathbb{R}^{N_r \times N_i}$, where $N_i$ is the number of input variables. Initially, all matrix entries are set to zero, and nonzero weights are assigned to rows corresponding to neurons in the cluster associated with each input.

For input $i$, the neurons in the $i^{th}$ cluster (of size $N_{\text{c}}$) receive nonzero weights, defined as:
\[
W_{\text{in}}[j, i] =
\begin{cases}
	w, & \text{if neuron } j \text{ belongs to cluster } i, \\
	0, & \text{otherwise},
\end{cases}
\]

here, $w \sim \text{uniform}(-1, 1)$, scaled by an input scaling factor that controls the magnitude of influence each input exerts on the reservoir dynamics. The sparsity of input connections is enforced using a random mask based on a predefined input connectivity parameter. The reservoir state update follows the standard ESNs formulation (Eq. \ref{eq: eq1}).

In the cluster-specific readout architecture, the reservoir is partitioned into $m$ clusters. For each cluster $i$ ($i=1,2,\dots,m$):
\begin{itemize}
	\item $u_i(t) \in \mathbb{R}$ represents the input component for cluster $i$, assuming $\mathbf{u}(t) \in \mathbb{R}^{n_{m}}$.
	\item $\mathbf{x}_i(t) \in \mathbb{R}^{N_c}$ denotes the state of cluster $i$ at time $t$, where $N_c$ is the number of neurons in the cluster.
\end{itemize}

The output for cluster $i$ is computed as:

\begin{equation}
	y_i(t) = W_{\text{out}}^{(i)} \mathbf{x}_i(t) ,
\end{equation} 
where $W_{\text{out}}^{(i)} \in \mathbb{R}^{1 \times (N_c+1)}$ is the readout weight vector for cluster $i$.

The overall system output is obtained by stacking the individual cluster outputs:

\begin{equation}
	\mathbf{y}(t) = \begin{bmatrix} y_1(t) \\ y_2(t) \\ \vdots \\ y_{n_{m}}(t) \end{bmatrix} \in \mathbb{R}^{n_{m}}.
\end{equation}
\bigskip

\textbf{Training via Ridge Regression:}

For each cluster $i$, we collect the extended reservoir states over $T$ time steps into the matrix:
\begin{equation}
	\mathbf{X}^{(i)} = \begin{bmatrix}
		\begin{array}{cccc}
			\mathbf{x}_i(1) & \mathbf{x}_i(2) & \cdots & \mathbf{x}_i(T)
		\end{array}
	\end{bmatrix} \in \mathbb{R}^{(N_c+1) \times T},
\end{equation}

where $u_i(t)$ represents the input time series for cluster $i$, and $\mathbf{x}_i(t)$ denotes the corresponding reservoir state at time step $t$.

Similarly, the target output matrix is defined as:
\begin{equation}
	\mathbf{y}^{(i)} = \begin{bmatrix}
		y_i(1) & y_i(2) & \cdots & y_i(T)
	\end{bmatrix} \in \mathbb{R}^{1 \times T}.
\end{equation}

The training process involves solving a ridge regression problem \cite{marquardt1975ridge} for each cluster $i$, formulated as:
\begin{equation}
	\min_{W_{\text{out}}^{(i)}} \; \left\| \mathbf{y}^{(i)} - W_{\text{out}}^{(i)} \mathbf{X}^{(i)} \right\|^2 + \lambda \left\| W_{\text{out}}^{(i)} \right\|^2.
\end{equation}

Here, the first term minimizes the squared error between the predicted and actual outputs. In contrast, the second term regularizes the weight matrix $W_{\text{out}}^{(i)}$ to prevent overfitting, which is controlled by the regularization parameter $\lambda$.

The closed-form solution for the optimal output weight matrix is given by:
\begin{equation}
	W_{\text{out}}^{(i)} = \mathbf{y}^{(i)}\,\mathbf{X}^{(i)\top} \left( \mathbf{X}^{(i)}\mathbf{X}^{(i)\top} + \lambda \mathbf{I} \right)^{-1}.
\end{equation}
This expression ensures numerical stability, especially when the number of training samples is small relative to the number of features. Once the reservoir is trained, the prediction for cluster $i$ at time $t$ is computed as:
\begin{equation}
	y_i(t) = W_{\text{out}}^{(i)}\mathbf{x}_i(t).
\end{equation}
This formulation highlights the role of both the input and the reservoir state in determining the cluster's output. 

The overall system output, aggregating predictions from all $m$ clusters, is given by:
\begin{equation}
	\hat{\mathbf{y}}(t) = \begin{bmatrix} y_1(t) \\ y_2(t) \\ \vdots \\ y_{n_{m}}(t) \end{bmatrix}.
\end{equation}

Another important consideration in the design of reservoir networks is the connection density, which directly influences the computational cost. For a general reservoir network, the connection density is defined as:
\begin{equation}
	\text{Density} = \frac{E}{N \cdot (N - 1)},
\end{equation}
where $E$ represents the total number of edges in the network and $N$ denotes the total number of nodes. The connection density is typically controlled by a parameter $p$, which represents the probability of forming a connection between any two nodes. This parameter can be adjusted based on specific computational and predictive requirements. In the case of a clustered reservoir, the total density can be computed by distinguishing between intra-cluster connections ($E_{intra}$) and inter-cluster connections ($E_{inter}$). The number of intra-cluster connections is calculated as:
\begin{equation}
	E_\text{intra} = m \cdot p_\text{in} \cdot N_c \cdot (N_c - 1),
\end{equation}
where  $p_{in}$  is the probability of intra-cluster connections. The inter-cluster connections ($E_{inter}$) are calculated as:
\begin{equation}
	E_\text{inter} = p_\text{out} \cdot \left[ N \cdot (N - 1) - E_{intra} \right],
\end{equation}
where $p_{out}$ is the probability of forming connections between different clusters. The total number of edges in the clustered reservoir is then:
\begin{equation}
	\text{Density} = \frac{E_{intra} +  E_{inter}}{N \cdot (N - 1)}.
\end{equation}

\begin{figure*}[!htbp] 
	\centering
	\includegraphics[width=1.0\linewidth]{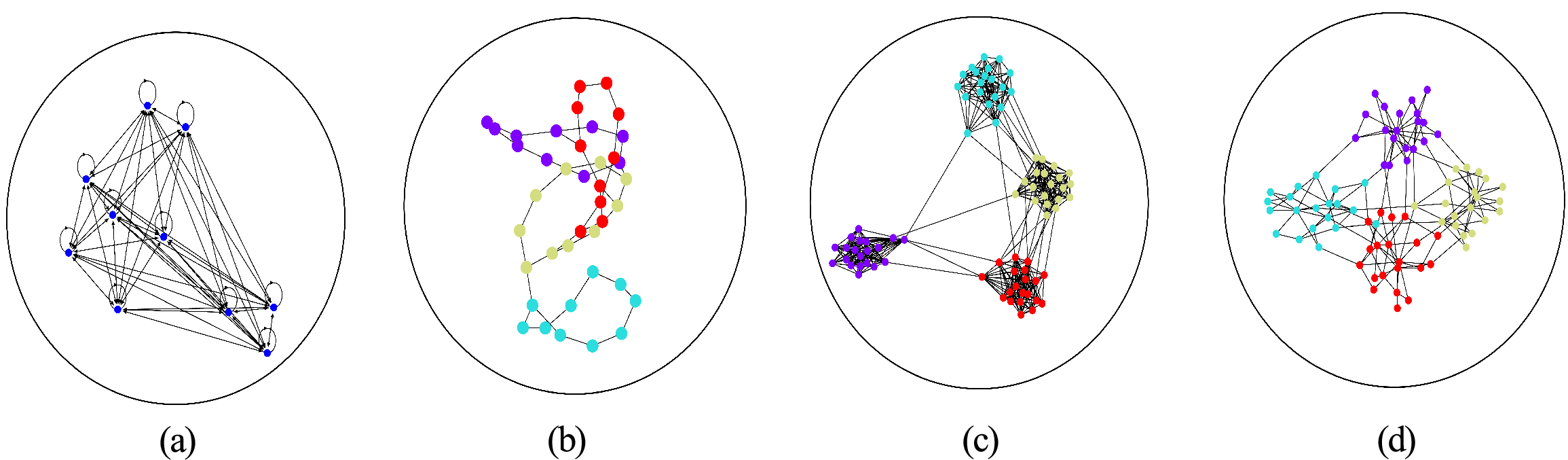}
	\caption{\label{topo}Illustration of different network topologies used in reservoir computing: (a) Conventional reservoir with randomly connected neurons, (b) clustered ring topology, (c) clustered ER network topology, (d) clustered BA topology. Each topology influences the reservoir's dynamical properties, affecting memory capacity and feature extraction in ESNs.}
\end{figure*}

\begin{figure*}[!htbp] 
	\centering
	\includegraphics[width=1.0\linewidth]{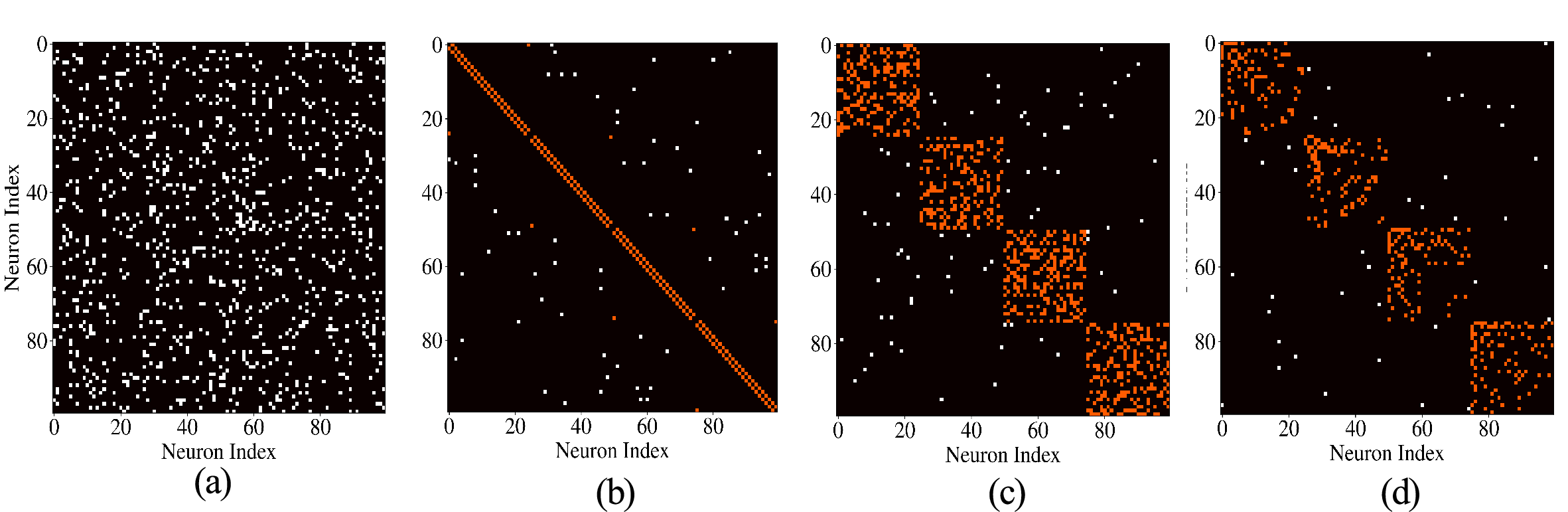}
	\caption{Structure matrix plots illustrating connectivity patterns in different ESN architectures: (a) Unclustered ESN, where white dots represent neuron-to-neuron connections, and black regions indicate the absence of connections; (b) Clustered ring topology, (c) Clustered ER network, and (d) Clustered SF network. In (b–d), white represents inter-cluster connections, black denotes no connections, and red highlights intra-cluster connections. These structural variations influence the reservoir's computational efficiency and ability to capture complex temporal dependencies in multivariate time series data.}%
	\label{struc}
\end{figure*}

This formulation allows a direct comparison between the density of conventional and clustered ESNs. For a conventional ESN, the density is solely determined by the parameter $p$. For instance, if $p = 0.3$, the density is approximately $0.3$ or $30\%$. In a clustered reservoir with $p_{in} = 0.3$ and $p_{out} = 0.01$ (to maintain sparsity between clusters), with $N = 80$, $m = 4$, and $N_c = 20$, the resulting density is approximately $0.071$ or $7.1\%$. This demonstrates that the clustered structure yields fewer connections, enhancing computational efficiency while maintaining the ability to capture multivariate dependencies.

Building on this foundational understanding of the CESNs for handling multiple inputs designated to specific clusters, we extend the framework to different reservoir architectures. Specifically, we focus on three topologies: ring networks, random networks generated using the ER model, and SF networks constructed using the BA model. Figure \ref{topo} illustrates these various network structures. In Fig. \ref{topo}(a), the conventional ESN is depicted with randomly distributed connections throughout the reservoir. Figure \ref{topo}(b) represents the clustered ring topology, where different colors indicate separate clusters, each processing one dimension of the multivariate time series. Figure \ref{topo}(c) shows the clustered ER network, while Fig. \ref{topo}(d) depicts the clustered SF network. In all three clustered topologies, intra-cluster connections are dense while inter-cluster connections remain sparse, allowing limited but meaningful interaction across clusters. This structure enhances the reservoir's ability to model complex dependencies among variables, as changes in one dimension propagate through inter-cluster connections.

\begin{table*}[!htbp] 
	\centering
	\renewcommand{\arraystretch}{1.5}
	\caption{Fixed hyperparameters used for performing prediction tasks}
	\label{table}
	\begin{tabular}{|c|c|c|c|}
		\hline
		Random ESNs & Ring network & Erdős-Renyi network & Scale-free network \\
		\hline
		\multicolumn{4}{|c|}{Leak rate ($\alpha$) = 1.0} \\
		\hline
		\multicolumn{4}{|c|}{Input scaling = 1.0} \\
		\hline
		\multicolumn{4}{|c|}{Input connectivity = 0.2} \\
		\hline
		\multicolumn{4}{|c|}{Regularization ($\lambda$) = $10^{-8}$} \\
		\hline
		\multirow{2}{*}{Reservoir connectivity ($p$) = 0.3} & $p_{\text{in}}$ = 1.0 & $p_{\text{in}}$  = 0.3 & Mean degree ($g$) = 5  \\
		\cline{2-4}
		& $p_{\text{out}}$ = 0.01 & $p_{\text{out}}$  = 0.01& $p_{\text{out}}$   = 0.01 \\
		\hline
	\end{tabular}
\end{table*}

To further elucidate the network structures, we present the corresponding structure matrices in Fig. \ref{struc}. In Fig. \ref{struc}(a), the structure matrix of a conventional random ESN is displayed, where black represents the absence of a connection and white denotes the presence of a connection. In Fig. \ref{struc}(b) and (c), the intra-cluster connections are shown in red, inter-cluster connections in white, and absent connections in black. For the ring topology, the bidirectional nature of connections is evident through paired structures within clusters, with sparse inter-cluster connections providing limited cross-cluster information flow. Figure \ref{struc}(c) illustrates the clustered ER network, where clusters are clearly visible, and the sparse inter-cluster connections follow the ER network. In Fig. \ref{struc}(d), the clustered SF network is shown, where a hub structure is prominent, controlled by the mean degree (\(g\)) within clusters and sparse inter-cluster connectivity. This diverse set of architectures offers flexibility to tailor the ESN to different types of multivariate time series.

\subsection{Parameter sensitivity pre-experiments}
\begin{figure}[!htbp] 
	\centering
	\includegraphics[width=1.0\linewidth]{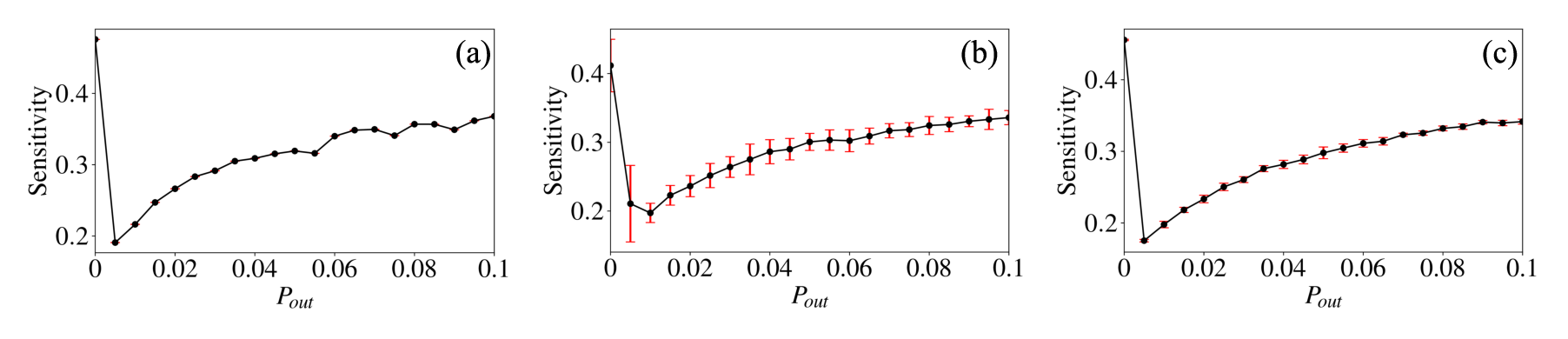}
	\caption{Sensitivity analysis of the network performance with respect to the inter-cluster connection probability $P_{out} \in [0, 0.1]$ and intra-cluster connectivity $P_{in}$ (specific to each network topology). Results are shown for three network configurations: (a) Ring topology ($P_{in}$ = 1; hence, no error bars are present), (b) ER topology ($P_{in} \in [0.01, 1.0]$), and (c) SF topology (mean degree $g \in [2, 10]$). Each data point represents the mean normalized root mean square error ($\overline{\text{NRMSE}}$) computed over 50 independent realizations, with error bars indicating the corresponding standard deviation.}
	\label{sensitivity} 
\end{figure}
The key hyperparameters affecting the stability and performance of the network are summarized in Table \ref{table}. Common parameters across all architectures include the leak rate ($\alpha$), spectral radius ($\rho$), ridge regularization ($\lambda$), input connectivity, and input scaling. Differences arise in the reservoir construction, where the clustered models introduce $p_{in}$ and $p_{out}$ to control intra- and inter-cluster connectivity, respectively. For SF networks, intra-cluster connectivity is regulated by the mean degree ($g$), while inter-cluster connectivity is governed by $p_{out}$.  

To systematically examine the influence of network connectivity, we design a controlled sensitivity experiment in which synthetic correlated white Gaussian noise is generated using a covariance matrix whose off-diagonal entries are fixed at a prescribed correlation coefficient 	$H =$ 0.8. This configuration ensures that the individual time series exhibit a strong linear interdependence. Ten correlated noisy time-series signals, each containing 5000 samples, are generated, resulting in $m =$ 10 clusters. The number of neurons per cluster is fixed at $N_c = 10$, yielding a total of $N = $ 100 reservoir nodes. Since each cluster processes one of the noisy time-series components, it is essential to determine the optimal inter-cluster connectivity probability $P_{out}$. To ensure consistency with the reservoir’s operational range and to promote stable, accurate predictions, the data are first normalized using min–max scaling to transform all input values into the range of [0, 1]. This normalization is mathematically expressed as:
	\begin{equation}
		x_{\text{scaled}} = \frac{x - \min(x)}{\max(x) - \min(x)},
		\label{scale}
	\end{equation}
	where $x$ is the original data value, and $x_{\text{scaled}}$ is the normalized output. This normalization mitigates the effects of varying data magnitudes and ensures uniform input dynamics. Following normalization, each topology undergoes a sensitivity analysis for $P_{out} \in [0,0.1]$, while varying $P_{in}$ (which depends on the specific network topology). The network is trained on 80 \% of the dataset and evaluated on the remaining 20 \%. Prediction accuracy is quantified using the normalized root-mean-squared error (NRMSE), a standard metric defined as:
	\begin{equation}
		\label{nrmse}
		\text{NRMSE} = \frac{\sqrt{\frac{1}{N} \sum_{t=1}^{N} \left( y_t - \hat{y}_t \right)^2}}{\sigma_y},
	\end{equation}
	where $y_t$ represents the true value, $\hat{y}_t$ denotes the predicted value, and $\sigma_y$ is the standard deviation of the actual output. For a comprehensive evaluation, the averaged NRMSE, denoted as ($\overline{\text{NRMSE}}$), is computed across all predicted variables to obtain an overall measure of model performance. For prediction, we set the hyperparameters as outlined in Table \ref{table}.

In the sensitivity analysis illustrated in Fig.~\ref{sensitivity}, for each value of $P_{out}$, the mean and standard deviation of the $\overline{\text{NRMSE}}$ are computed by averaging over all intra-cluster connectivity values ($P_{in}$) and 50 independent realizations, thereby quantifying the network's sensitivity to inter-cluster coupling. For the ring topology (Fig.~\ref{sensitivity}(a)), $P_{in} = 1$ since all neurons are connected to their immediate neighbors; consequently, no error bars are observed. It is evident that $P_{out} = 0.01$ yields the lowest prediction error across realizations, while the error increases gradually with larger $P_{out}$ values. For the ER network (Fig.~\ref{sensitivity}(b)), $P_{in}$ is varied within the range [0.01, 0.5] for each $P_{out}$ across 50 realizations. The configuration $P_{out} = 0.01$ performs better than both the fully independent clusters and highly connected cases ($P_{out}>0.05$). A similar pattern is observed for the SF network (Fig.~\ref{sensitivity}(c)), where the mean degree $g$ represents the intra-cluster connectivity ($P_{in}$ and is varied within the range $g \in [2,10]$. Overall, the results suggest that $P_{out}$ should be selected to ensure that the clusters are neither completely independent nor excessively interconnected, thereby preserving the network’s modular structure and enhancing predictive stability.

\subsection{Comprehensive performance analysis of reservoir topologies}
\begin{table}[htbp]
	\centering
		\renewcommand{\arraystretch}{1.3}
		\setlength{\tabcolsep}{6pt}
		\resizebox{\textwidth}{!}{%
			\begin{tabular}{c|ccc|ccc|ccc}
				\toprule
				\multirow{2}{*}{\textbf{Neurons per cluster}} 
				& \multicolumn{3}{c|}{\textbf{Training Time (seconds)}} 
				& \multicolumn{3}{c|}{\textbf{Memory Usage (MB)}} 
				& \multicolumn{3}{c}{\textbf{$\overline{\text{NRMSE}}$}} \\
				\cline{2-10}
				& Ring & ER & SF 
				& Ring & ER & SF 
				& Ring & ER & SF \\
				\midrule
				10  & \textbf{0.28 $\pm$ 0.02} & \textbf{0.28 $\pm$ 0.01} & 0.31 $\pm$ 0.02 
				& \textbf{3.92 $\pm$ 8.46e-04} & 4.00 $\pm$ 4.48e-04 & 4.01 $\pm$ 0.02 
				& 0.20 $\pm$ 0.09 & 0.20 $\pm$ 0.11 & \textbf{0.18 $\pm$ 0.08} \\
				
				20  & 0.56 $\pm$ 0.02 & 0.58 $\pm$ 0.02 & \textbf{0.53 $\pm$ 0.05}
				& \textbf{8.00 $\pm$ 1.58e-04} & 8.32 $\pm$ 1.54e-04 & 8.17 $\pm$ 0.02 
				& 0.084 $\pm$ 0.04 & \textbf{0.08 $\pm$ 0.05} & \textbf{0.06 $\pm$ 0.02} \\
				
				40  & 1.45 $\pm$ 0.05 & 1.58 $\pm$ 0.05 & \textbf{0.99 $\pm$ 0.07}
				& \textbf{16.64 $\pm$ 2.25e-04} & 17.92 $\pm$ 2.54e-04 & 17.09 $\pm$ 0.05 
				& 0.05 $\pm$ 0.01 & \textbf{0.04 $\pm$ 0.01} & \textbf{0.04 $\pm$ 0.01} \\
				
				60  & 3.07 $\pm$ 0.12 & 3.54 $\pm$ 0.13 & \textbf{1.82 $\pm$ 0.11}
				& \textbf{25.94 $\pm$ 3.12e-04} & 28.82 $\pm$ 3.17e-04 & 26.74 $\pm$ 0.15 
				& 0.05 $\pm$ 0.04 & \textbf{0.04 $\pm$ 0.01} & \textbf{0.04 $\pm$ 0.005} \\
				
				80  & 5.23 $\pm$ 0.25 & 5.79 $\pm$ 0.28 & \textbf{2.82 $\pm$ 0.15}
				& \textbf{35.89 $\pm$ 3.52e-04} & 41.02 $\pm$ 3.41e-04 & 37.08 $\pm$ 0.20 
				& 0.03 $\pm$ 0.03 & \textbf{0.02 $\pm$ 0.03} & \textbf{0.02 $\pm$ 0.003} \\
				
				100 & 8.05 $\pm$ 0.30 & 8.31 $\pm$ 0.30 & \textbf{4.00 $\pm$ 0.16}
				& \textbf{46.50 $\pm$ 2.76e-04} & 54.50 $\pm$ 2.64e-04 & 48.02 $\pm$ 0.16 
				& 0.03 $\pm$ 0.04 & \textbf{0.02 $\pm$ 0.03} & \textbf{0.02 $\pm$ 0.004} \\
				\bottomrule
				
			\end{tabular}
		} 
	\caption{Comparison of training time, memory usage, and mean normalized root mean square error ($\overline{\text{NRMSE}}$) for different network topologies and varying numbers of neurons per cluster ($N_c$). The best-performing values in each category are highlighted in \textbf{bold}}.
	\label{perf_table}
\end{table}

In order to understand the performance of reservoirs with the proposed topologies, such as ring, ER, and SF networks, we perform a detailed analysis considering both predictive and non-predictive indicators. These include training time, memory usage, and prediction accuracy, as measured by NRMSE. To compute memory usage, we employ the \texttt{tracemalloc} Python package to monitor memory consumption from reservoir initialization through prediction. The same correlated-noise task described earlier is undertaken, varying the number of neurons per cluster ($N_c$), as presented in Table~\ref{perf_table}. As observed from Table~\ref{perf_table}, the SF network exhibits the lowest training time among the three topologies, while the ring topology consumes the least memory owing to its structured and highly regular connectivity. In terms of prediction performance, the SF network consistently achieves the lowest NRMSE values, outperforming both the clustered ring and ER topologies across all values of $N_c$.	The ER network, on the other hand, offers a balanced trade-off between computational cost and predictive accuracy--its performance becomes comparable to that of the SF topology for $N_c > 10$, but with slightly higher memory consumption and training time. This comparative analysis highlights the flexibility of the proposed CESN framework, demonstrating that the choice of topology can be adapted to specific task requirements. For instance, the SF topology is preferable for achieving high prediction accuracy, whereas the ring network is advantageous for applications requiring low memory overhead. The ER network offers a middle ground, striking a balance between accuracy, training efficiency, and resource usage. Such tunability enables practitioners to select the most suitable reservoir topology depending on dataset size, noise level, and computational constraints.
\begin{figure}[!htbp] 
	\centering
	\includegraphics[width=0.5\linewidth]{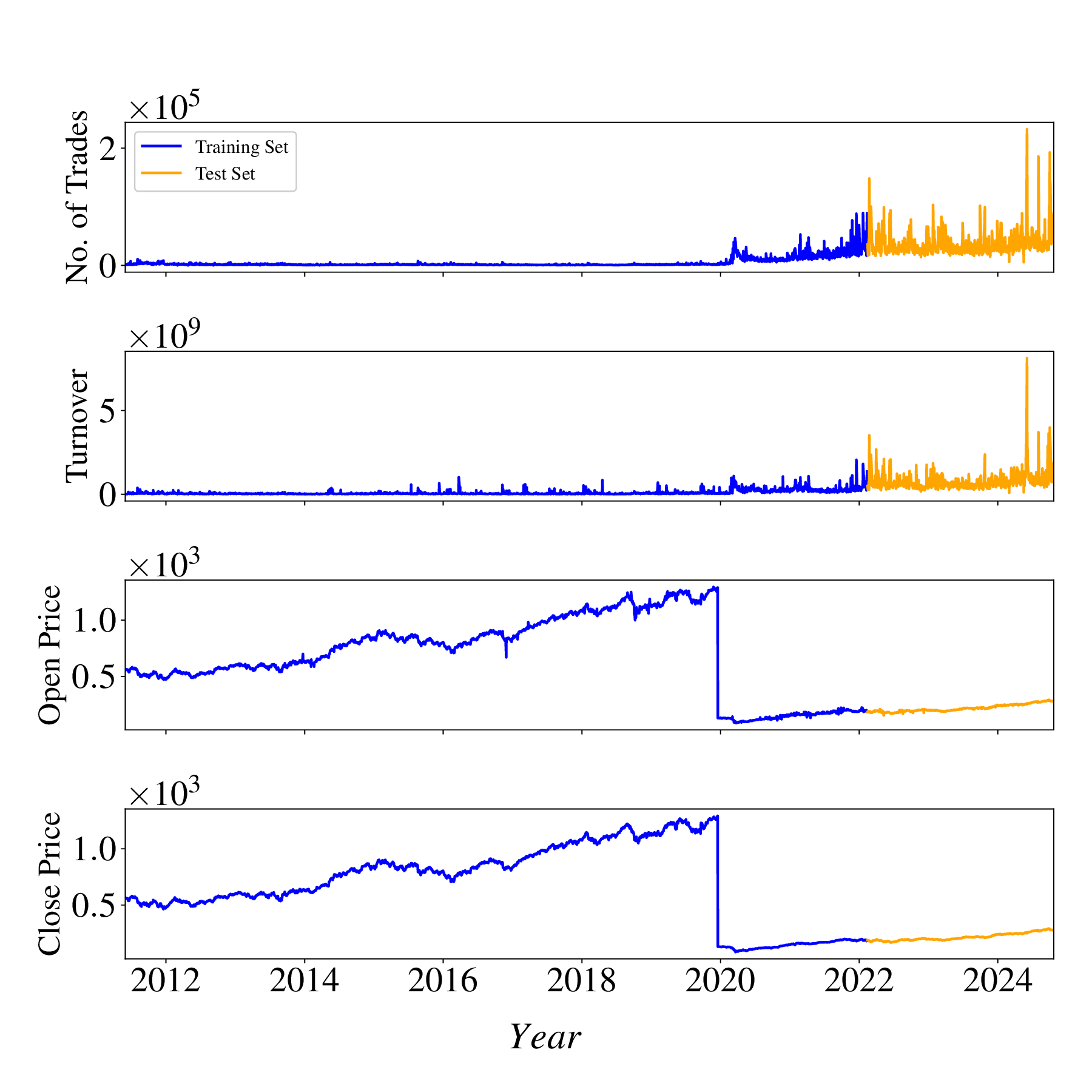}
	\caption{Stock market multivariate time series data obtained from NSE, India. The selected parameters for prediction include the number of trades, turnover, and the open and close prices of NIFTYBEES. The dataset is divided into an 80\% training set (blue) and a 20\% test set (orange) for model evaluation.}
	\label{stock} 
\end{figure}

\begin{figure*}[!htbp] 
	\centering
	\includegraphics[width=0.9\linewidth]{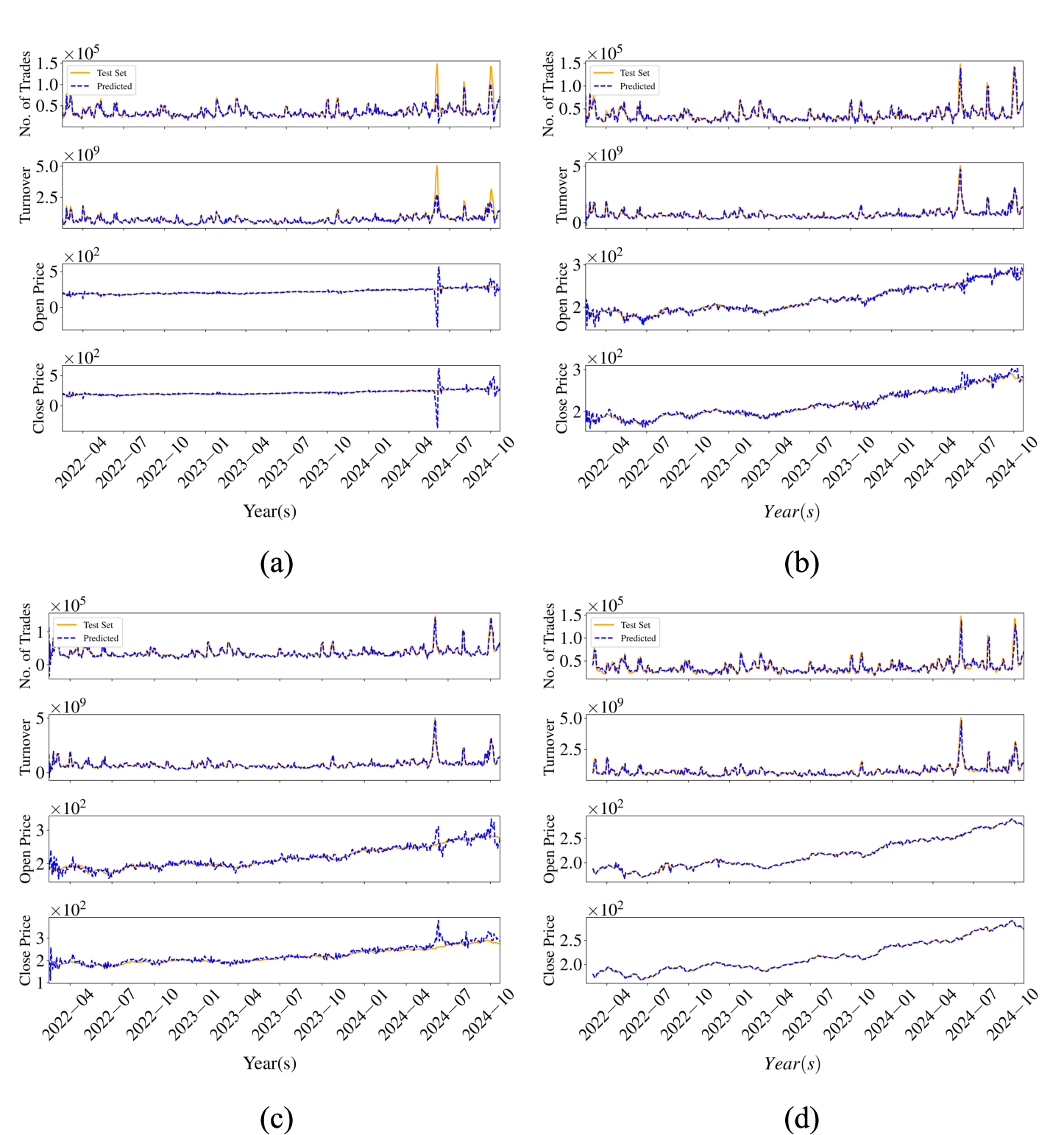}
	\caption{Prediction of the test time series for NIFTYBEES (NSE, India) using different ESN models: (a) Conventional ESN, (b) clustered Ring ESN, (c) clustered SF ESN, and (d) clustered ER ESN. The actual test data is represented in orange, while the model predictions are shown in blue. }%
	\label{st_pred} 
\end{figure*}

\section{Performance of CESNs}
\label{sec4}
\subsection{Prediction of complex real-world data set}
\label{sec41}
The primary objective of this work is to predict multivariate time series, with a focus on the challenging task of forecasting stock market data, which is inherently noisy and unpredictable. Figure \ref{stock} presents the multivariate time series derived from the Indian NSE data, represented by the NIFTYBEES exchange-traded fund \cite{nse_all_reports}. NIFTYBEES closely tracks the performance of the NIFTY 50 index. This dataset provides a comprehensive view of market dynamics and is widely used for analyzing financial trends, volatility, and the impact of macroeconomic factors on equity markets. The dataset comprises four variables: the number of trades, turnover, open price, and close price, spanning the period from June 1, 2011, to October 24, 2024. As observed, the number of trades and turnover exhibit values on the order of $10^{9}$, requiring data preprocessing to align with the dynamic range of the reservoir \cite{jaeger2002adaptive}.

Initially, the data are normalized using min–max scaling as described in Eq.~(\ref{scale}), and a one-dimensional Gaussian filter with a standard deviation of $\sigma = 1$ is applied to suppress high-frequency noise while preserving the temporal structure of the signals. This smoothing enhances the reservoir’s ability to capture meaningful temporal correlations, thereby improving the overall stability of predictions. For model training and testing, the hyperparameters are configured as listed in Table~\ref{table}. The model performance is evaluated using the NRMSE defined in Eq.~(\ref{nrmse}), and the averaged NRMSE ($\overline{\mathrm{NRMSE}}$) is computed across all four predicted variables to provide a holistic measure of predictive accuracy.

\begin{figure*}[!htbp] 
	\centering
	\includegraphics[width=1.0\linewidth]{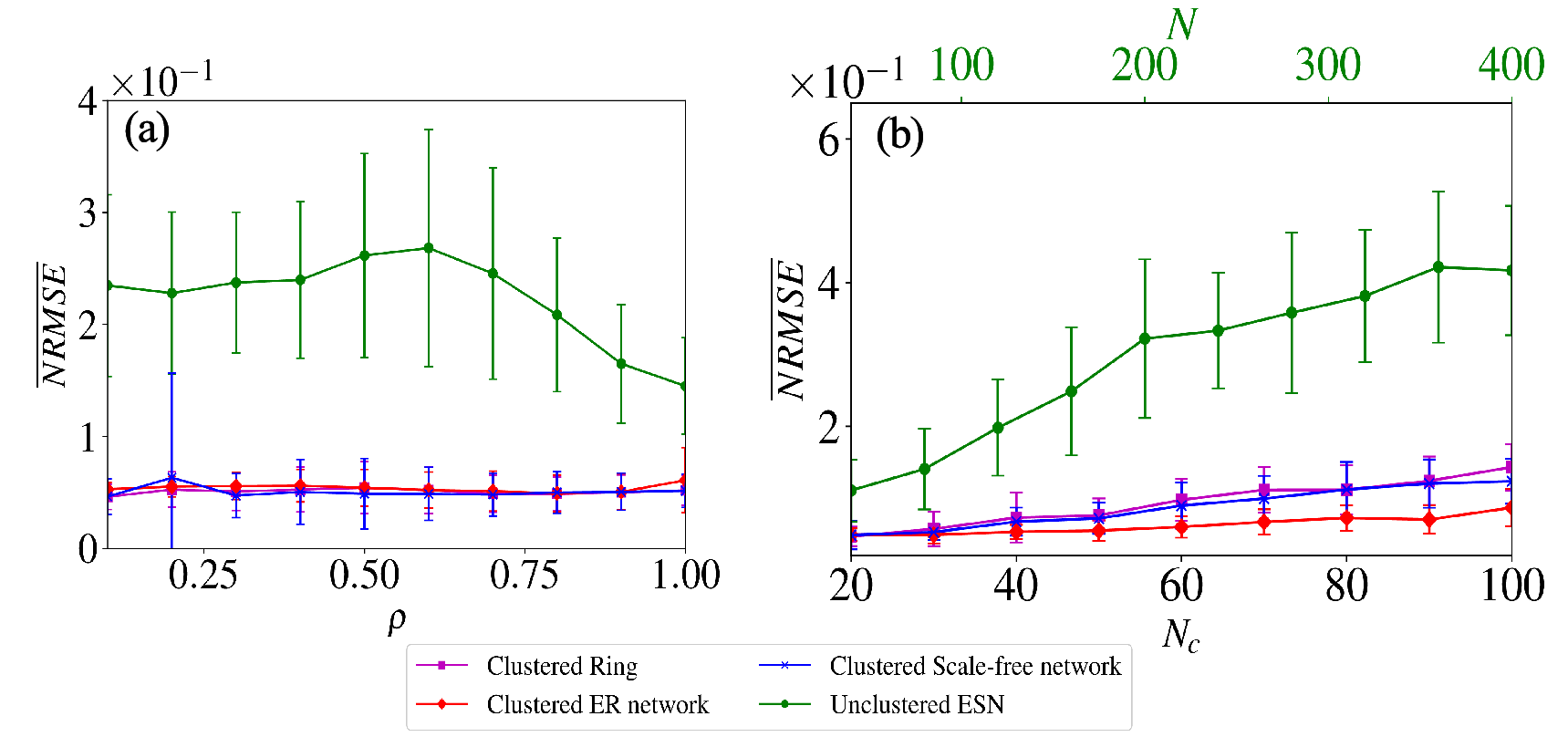}
	\caption{Averaged NRMSE ($\overline{\text{NRMSE}}$) for stock market time series data computed over 50 independent realizations, with error bars representing the standard deviation across trials. The analysis is presented for two configurations: (a) Variation of the spectral radius ($\rho$) while keeping the total number of nodes fixed at $N$ = 80 and the number of nodes per cluster at $N_c$ = 20;	(b) Comparison between unclustered ESNs (depicted by the green $\times$ curve) with varying total number of nodes $N$, and clustered ESNs (with number of clusters $m$ = 4) where the number of nodes per cluster $N_c$ is varied, under a fixed spectral radius $\rho$ = 0.9. }%
	\label{st_nrmse}
\end{figure*}

\begin{table}[!htbp]
	\centering
	\renewcommand{\arraystretch}{1.2} 
	\caption{Average NRMSE ($\overline{\text{NRMSE}}$) errors for stock market time series prediction using an 80\%-20\% train-test split.}
	\label{nrmse_stock}
	\begin{tabular}{|l|c|c|}
		\hline
		\textbf{Method} & Reservoir size & \textbf{$\overline{\text{NRMSE}}$} \\ \hline \hline
		Conventional ESN \cite{jaeger2002adaptive}    & $N = 100$   & $0.1966$ \\ \hline
		AEESN  \cite{xu2016adaptive}     & $N = 100$        & $0.08818$ \\ \hline
		DeepESN \cite{malik2016multilayered} & $N = [20,20]$ & $0.1033$ \\ \hline
		BrESN\cite{yao2019broad}  & $N = [20,20,20]$  & $0.08055$ \\ \hline
		Clustered Ring ESN   & $N_c = 20$    & \textbf{\boldmath $0.057287$} \\ \hline
		Clustered ER network & $N_c = 20$ & \textbf{\boldmath $ 0.049208$} \\ \hline
		Clustered SF network & $N_c = 20$              & \textbf{\boldmath $0.052794$} \\ \hline
	\end{tabular}
\end{table}

In this study, we compare the performance of our CESNs with various conventional ESNs, as well as advanced models such as AEESN, DeepSESN, and BrESN. For the conventional ESN and AEESN models, the reservoir size is fixed at $ N = 100$. In the CESN framework, we employ four clusters ($m$ = 4), each containing $N_c$ = 20 nodes. Table \ref{nrmse_stock} presents the $\overline{\text{NRMSE}}$ values, demonstrating that clustered networks consistently outperform conventional and enhanced ESN variants by significantly reducing prediction error across all metrics. Among these, the clustered ER network achieves the lowest $\overline{\text{NRMSE}}$, underscoring its superior predictive accuracy in handling complex time series data.
\begin{figure*}[!htbp] 
	\centering
	\includegraphics[width=1.0\linewidth]{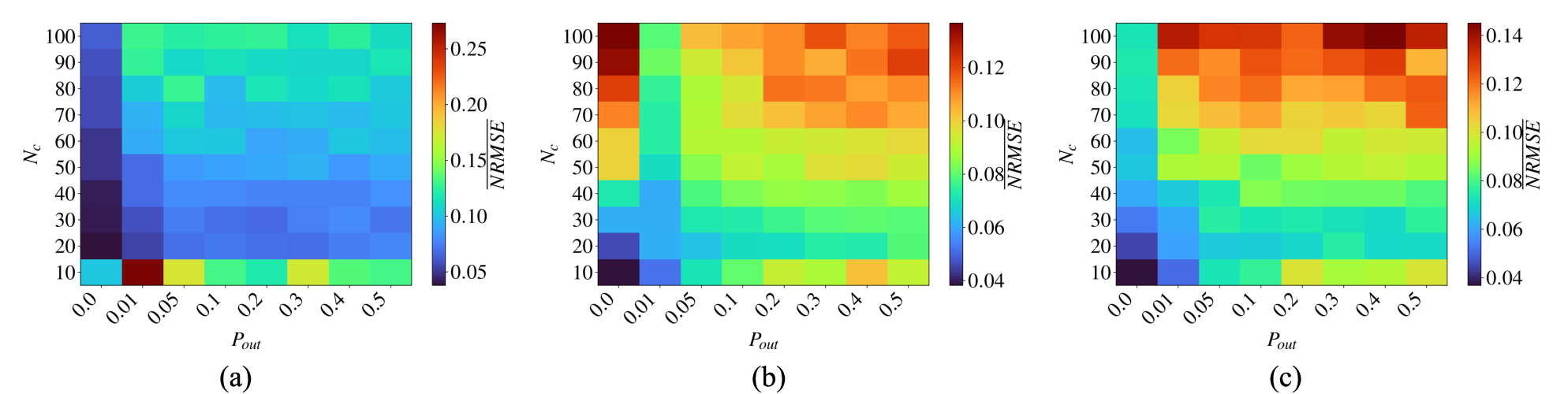}
	\caption{Two-parameter analysis of output connectivity probability ($P_{\text{out}}$) versus the number of nodes per cluster ($N_c$) for predicting stock market time series data. The color bar represents the averaged NRMSE ($\overline{\text{NRMSE}}$), computed over 50 independent realizations for each parameter combination. All other hyperparameters are fixed as specified in Table~\ref{table}. This analysis is performed across three different clustered network topologies: (a) Clustered Ring, (b) Clustered ER, and (c) Clustered SF networks. The goal is to identify optimal configurations for prediction accuracy and to highlight the sensitivity and limitations of $P_{\text{out}}$ under each topology. } %
	\label{st_bar}
\end{figure*}

Figure \ref{st_pred} provides a visual comparison between the predicted outputs (blue dashed line) and the actual test set (orange solid line) for each algorithm. As shown in Fig. \ref{st_pred}(a), the conventional ESN captures the patterns in the number of trades and turnover reasonably well but struggles with the open and close prices. The clustered ring (Fig. \ref{st_pred}(b)) and SF networks (Fig. \ref{st_pred}(c)) mitigate these errors, particularly for the open and close prices, resulting in lower $\overline{\text{NRMSE}}$. However, the SF network underperforms in predicting the number of trades and turnover. In contrast, the clustered ER network (Fig. \ref{st_pred}(d)) demonstrates superior accuracy in capturing all four variables compared to previously proposed algorithms, highlighting its effectiveness in modeling complex and noisy time series.

To further assess the robustness and stability of the proposed algorithms, we analyze their performance as we vary $\rho$, and $N$ for the conventional ESN and $N_c$ for the clustered models. For each parameter setting, we conduct $50$ independent trials and average the results to ensure statistical reliability.

Figure \ref{st_nrmse}(a) shows the impact of spectral radius variation with $N = 80$ and $N_c = 20$. At lower values of $\rho$, the conventional ESN exhibits poor performance, with a sharp reduction in error only when $\rho > 0.8$. Conversely, the clustered algorithms remain stable across the range $\rho \in [0.1, 1]$, with the clustered ring and ER networks demonstrating particularly robust performance.

\begin{figure}[!htbp] 
	\centering
	\includegraphics[width=0.5\linewidth]{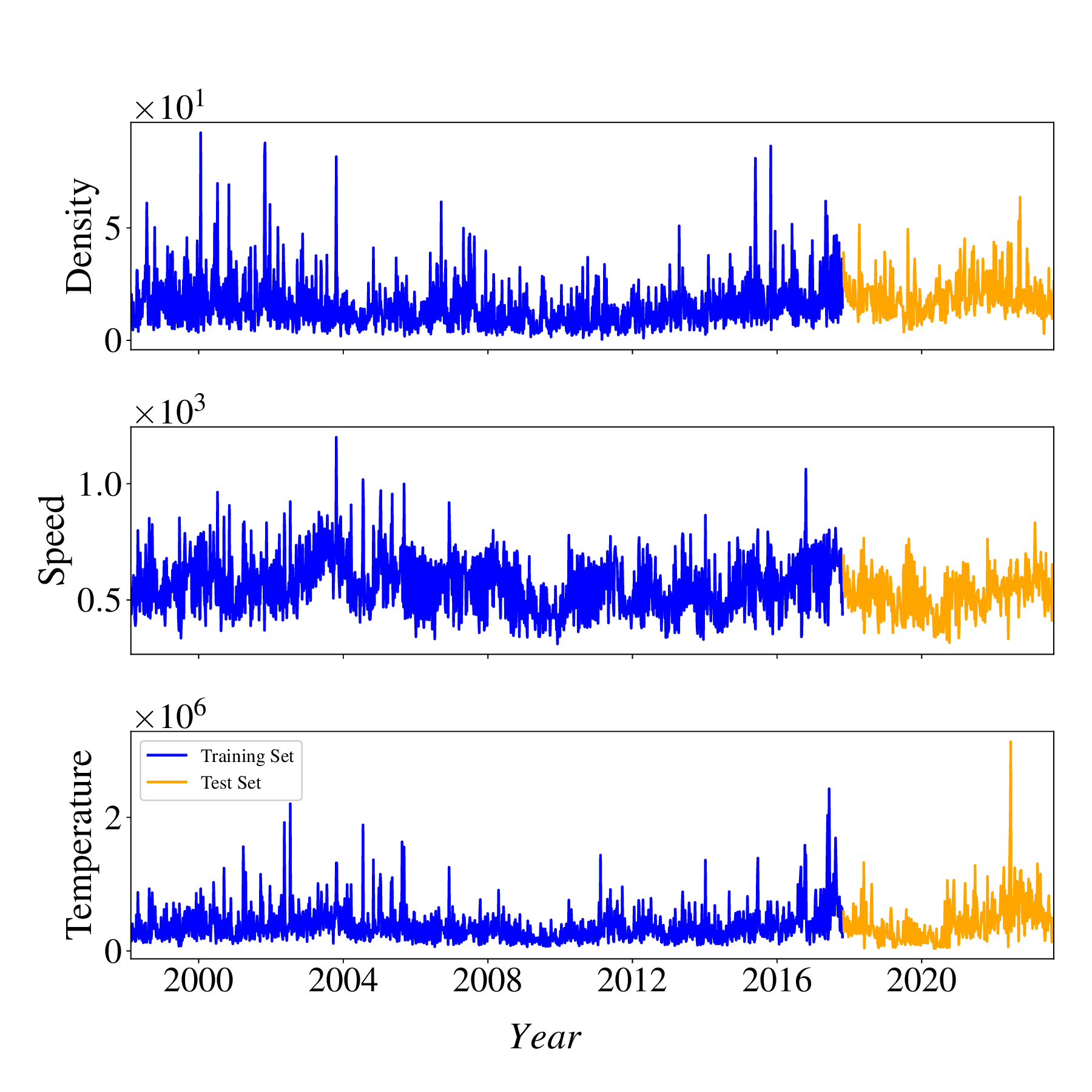}
	\caption{Solar wind data obtained from NOAA, covering the period from 1998 to 2024. The selected parameters for analysis include solar wind density, speed, and temperature. The dataset is divided into an 80\% training set (blue) and a 20\% test set (orange) for predictive modeling.}%
	\label{sun}
\end{figure}
\begin{figure*}[!htbp] 
	\centering
	\includegraphics[width=0.9\linewidth]{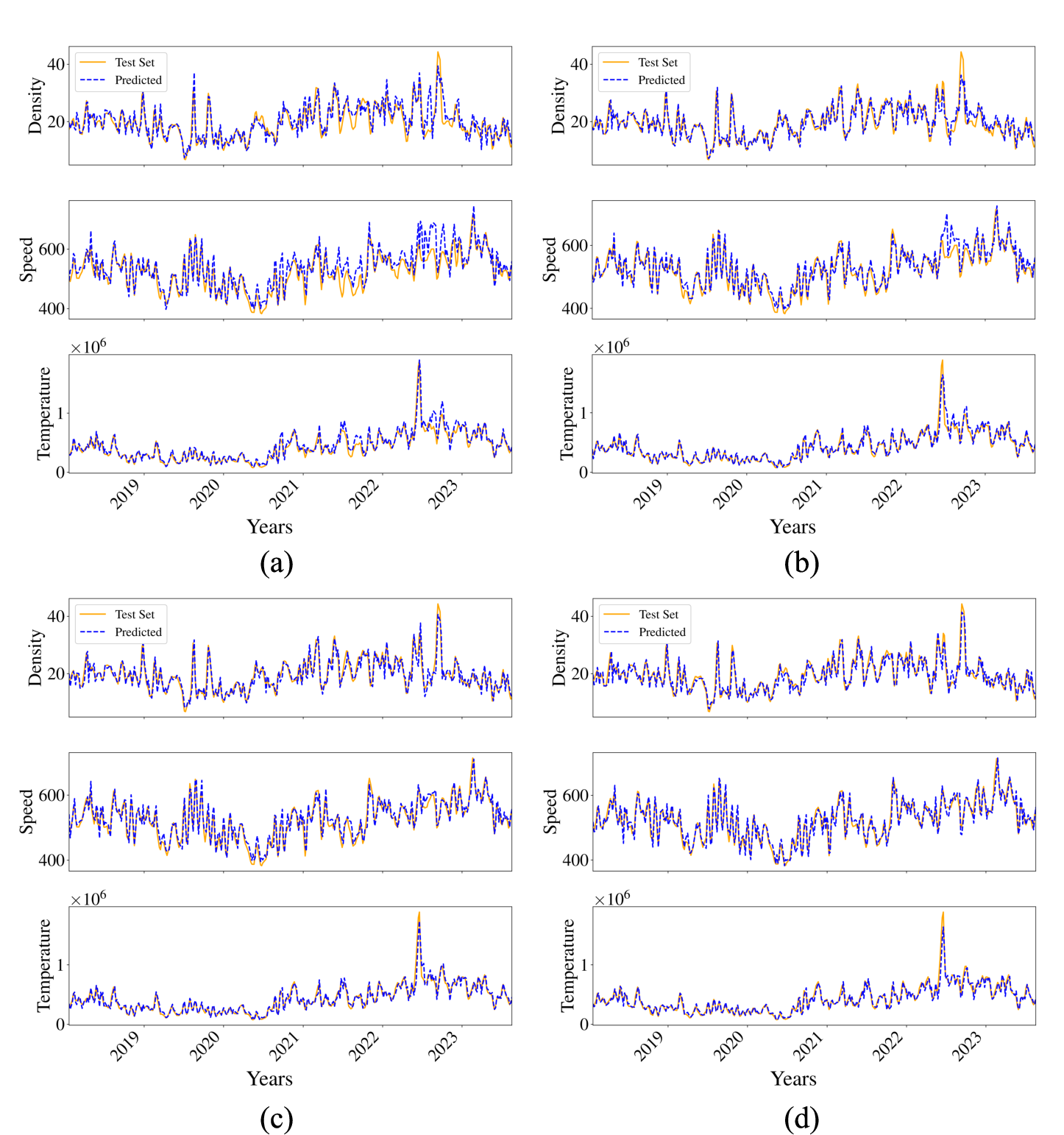}
	\caption{Prediction of the test time series for solar data using different ESN models: (a) Conventional ESN, (b) clustered Ring ESN, (c) clustered SF ESN, and (d) clustered ER ESN. The actual test data is represented in orange, while the model predictions are shown in blue dashed lines. }%
	\label{sun_pred} 
\end{figure*}
Figure \ref{st_nrmse}(b) presents the effect of increasing the reservoir size. While the error for the conventional ESN increases with larger $N$, the clustered algorithms maintain lower and more stable error levels as $N_c$ increases. This indicates that mapping input variables to specific clusters enhances the network's ability to capture intricate temporal dependencies, particularly for the ER and SF topologies. Notably, the performance of the clustered ring network deteriorates at higher $N_c$, suggesting its limitations in capturing long-term dependencies.
\begin{figure*}[!htbp] 
	\centering
	\includegraphics[width=1.0\linewidth]{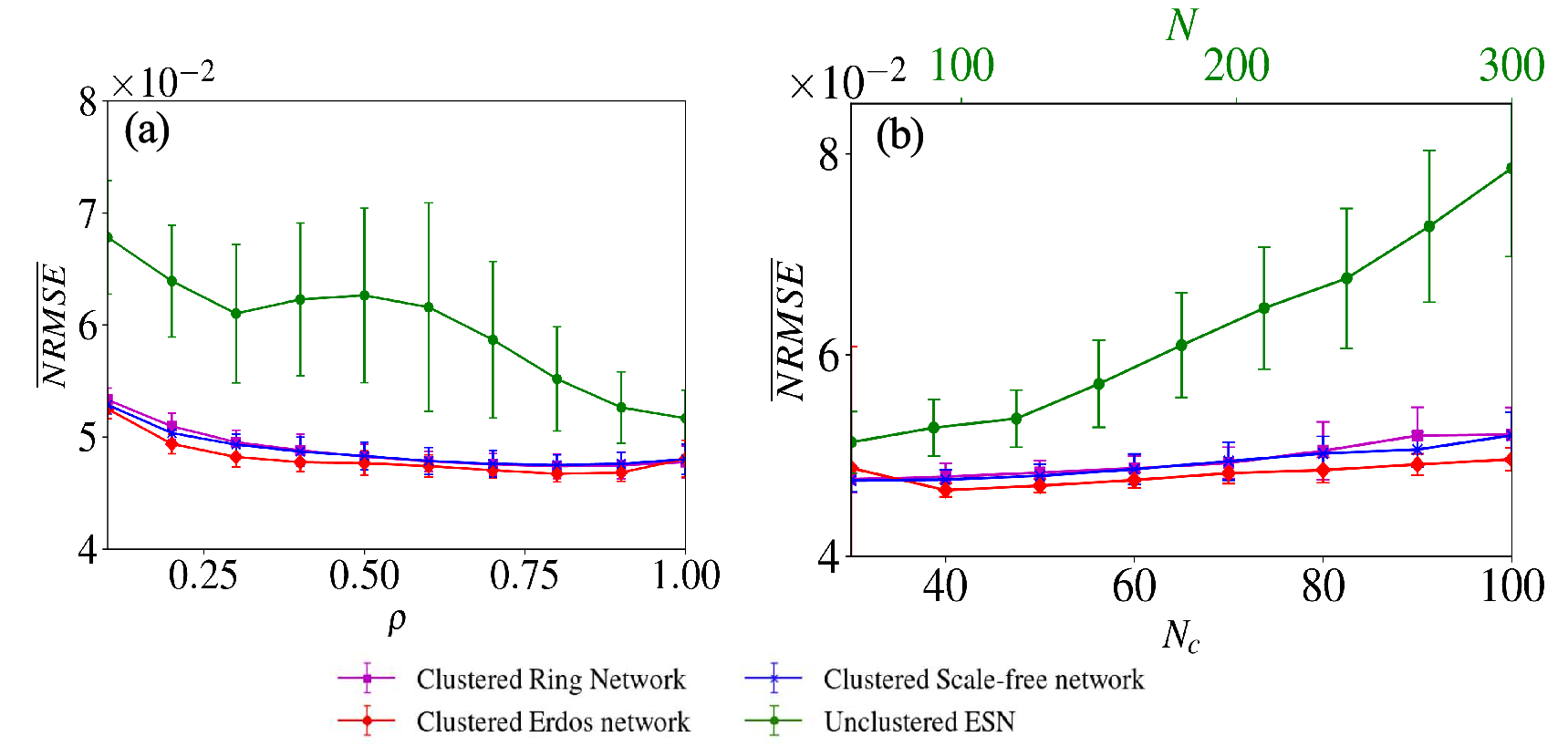}
	\caption{Averaged NRMSE ($\overline{\text{NRMSE}}$) plots for the solar wind time series dataset, evaluated over 50 independent realizations, under different reservoir configurations in CESNs. Panel (a) shows the effect of varying the spectral radius ($\rho$) while keeping the total number of reservoir nodes fixed at $N = 100$ and the number of neurons per cluster at $N_c = 25$. Panel (b) compares the performance of unclustered ESNs (indicated by green $\times$ markers) and clustered ESNs by varying the total number of nodes ($N$) and the number of neurons per cluster ($N_c$), respectively, with the spectral radius fixed at $\rho = 0.9$.} 
	\label{nrm}%
\end{figure*}

Since each cluster in the reservoir learns independently, it is crucial to investigate whether establishing inter-cluster connections improves performance. Specifically, we aim to determine the threshold at which clusters can function autonomously and when they become reliant on inter-cluster communication. To explore this, we vary the inter-cluster connection probability $P_{\text{out}} \in [0.0, 0.5]$, ranging from complete independence to maximal interconnectivity, across three different topologies. For each combination of $N_c$ and $P_{\text{out}}$, the analysis is repeated over 50 independent realizations, as shown in Fig.~\ref{st_bar}. In the case of the clustered ring topology (Fig.~\ref{st_bar}(a)), the model performs well even under full independence ($P_{\text{out}}$ = 0). Interestingly, a very weak inter-cluster coupling ($P_{\text{out}}$ = 0.01) yields comparable performance, particularly when the number of nodes per cluster $N_c \leq 50$. However, for larger cluster sizes ($N_c >$ 50), the performance under weak coupling slightly degrades compared to the fully independent configuration. As $P_{\text{out}}$ increases beyond 0.05, the performance stabilizes and remains competitive. This can be attributed to the gradual randomization of the structured ring topology with increasing inter-cluster links, which benefits learning in smaller clusters. For the clustered ER network (Fig.~\ref{st_bar}(b)), optimal predictive performance is observed for $P_{\text{out}}$ = 0.01 and $P_{\text{out}}$ = 0.05. This indicates that weak inter-cluster connectivity in random networks can enhance prediction accuracy compared to both completely isolated and highly interconnected configurations.	In the case of the clustered SF network, weakly dependent clusters perform well up to $N_c$ = 40, beyond which fully independent clusters tend to outperform the weakly connected ones. This outcome highlights the topology-dependent nature of inter-cluster dynamics and underscores the importance of tailoring inter-cluster connectivity to the network structure. 

Overall, these findings reveal that while independent clusters may dominate in certain scenarios, incorporating weak inter-cluster links can enhance learning efficiency, especially in random topologies, without the need for deploying multiple separate ESNs. This clustered architecture not only improves prediction performance over traditional ESNs but also reduces model complexity by limiting the number of hyperparameters.

\begin{figure*}[!htbp] 
	\centering
	\includegraphics[width=1.0\linewidth]{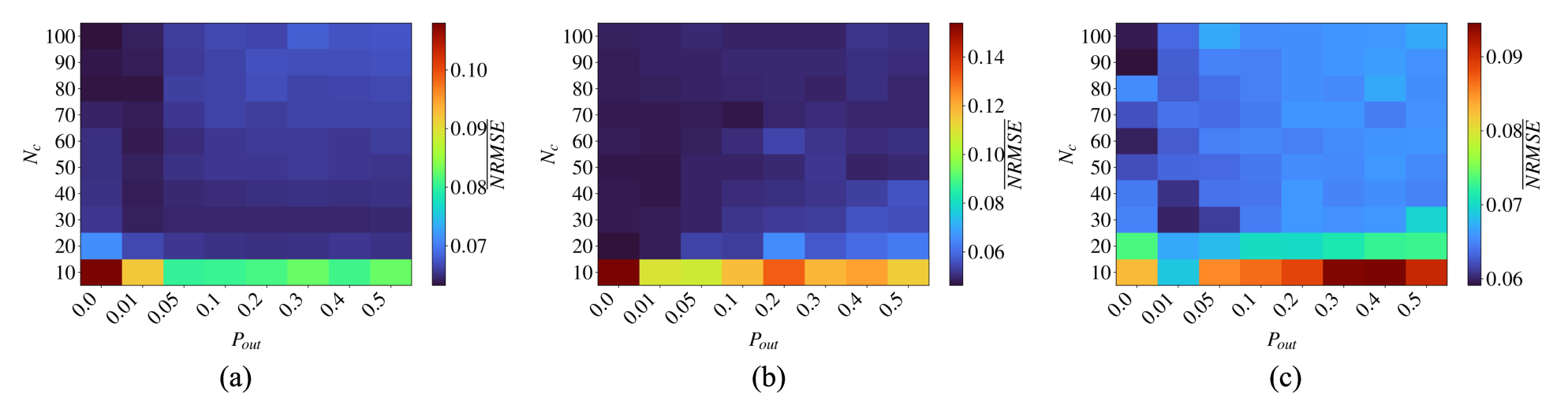}
	\caption{Two-parameter analysis of $P_{\text{out}}$ versus $N_c$ to identify optimal configurations for accurate prediction and to assess the limitations of inter-cluster connectivity ($P_{\text{out}}$), with the color bar representing the averaged NRMSE ($\overline{\text{NRMSE}}$) for the solar wind time series dataset. All other parameters are fixed as specified in Table~\ref{table}. Each parameter combination is evaluated over 50 independent realizations. Results are shown for three clustered network topologies: (a) ring, (b) ER, and (c) SF networks. }%
	\label{solar_bar}
\end{figure*}

To evaluate the model's generalization capabilities, we extend the analysis to additional noisy real-world time series datasets, such as solar wind data, as depicted in the Figure. \ref{sun}. The dataset is divided into an 80\% training set (blue) and a 20\% test set (orange).

\begin{itemize}
	\item \textbf{Solar Wind Data (Fig. \ref{sun}):} This dataset, obtained from the NOAA DISCOVR satellite \cite{noaa_solar_wind}, includes physical parameters of the solar wind (temperature, speed, and density) spanning the period from 1998 to 2024.
\end{itemize}

For the solar wind dataset, the clustered ER network yields the best predictive performance, as detailed in Table~\ref{nrmse_sun}. To visually illustrate this performance improvement, Fig. \ref {sun_pred} presents the predicted signals (blue dashed lines) overlaid on the original test data (orange solid lines). As shown in Fig.\ref{sun_pred}(a), the conventional ESN struggles to accurately capture the temporal dynamics, particularly for the $speed$ variable, leading to notable discrepancies between the prediction and ground truth. In contrast, the clustered ring topology (Fig.\ref{sun_pred}(b)) shows improvement in tracking the overall trends, yet it still falls short in precisely capturing finer temporal patterns. These limitations are significantly mitigated in the SF and ER topologies, as illustrated in Fig.\ref{sun_pred}(c) and (d), respectively. Among these, the clustered ER network exhibits the most consistent performance, accurately tracking the behavior of all three variables, thereby demonstrating its strength in handling complex, noisy multivariate signals such as solar wind data.

Figures \ref{nrm}(a) and \ref{nrm}(b) show that clustered ER and SF networks remain stable across varying $\rho$ and $N_c$ values, outperforming other conventional ESNs. This suggests that the clustered architecture provides a more robust framework for handling complex and highly noisy real-world time series data.

Next, we examine how inter-cluster connectivity influences prediction performance as a function of cluster size ($N_c$), with the results presented in Fig. \ref{solar_bar}. In the case of the ring topology (Fig.\ref{solar_bar}(a)), a weak inter-cluster connection with $P_{\text{out}} = 0.01$ consistently yields the best performance across all values of $N_c$, outperforming both fully independent clusters and other interconnection levels. For the clustered ER network (Fig.\ref{solar_bar}(b)), both fully independent clusters and weakly connected configurations ($P_{\text{out}} = 0.01$ and $0.05$) demonstrate nearly equivalent predictive accuracy, suggesting that minimal interdependence is sufficient for optimal performance in random networks. In contrast, the clustered SF network (Fig.\ref{solar_bar}(c)) shows a more nuanced behavior: fully independent clusters achieve the lowest averaged NRMSE ($\overline{\text{NRMSE}}$) when $N_c > 90$, while the best performance for weak interdependence ($P_{\text{out}} = 0.01$) is observed in the range $N_c \in [30, 40]$. All other interconnection probabilities result in higher prediction errors compared to these two optimal scenarios. These findings highlight the importance of adjusting both cluster size and inter-cluster connectivity according to the underlying network topology to achieve robust time series prediction.
\begin{table}[!htbp]
	\centering
	\renewcommand{\arraystretch}{1.2} 
	\caption{Average NRMSE ($\overline{\text{NRMSE}}$) errors for solar multivariate time series prediction using an 80\%-20\% train-test split.}
	\label{nrmse_sun}
	\begin{tabular}{|l|c|c|}
		\hline
		\textbf{Method} & Reservoir size & \textbf{$\overline{\text{NRMSE}}$} \\ \hline \hline
		Conventional ESNs \cite{jaeger2002adaptive}    & $N = 300$               & \textbf{\boldmath $0.0752$}\\ \hline
		AEESNs  \cite{xu2016adaptive}              & $N = 100$               & $0.0778$ \\ \hline
		DeepESNs \cite{malik2016multilayered}             & $N = [100,100]$  & $0.1042$ \\ \hline
		BrESNs\cite{yao2019broad}                & $N = [100,100,100]$ & $0.0763$ \\ \hline
		Clustered Ring ESNs   & $N_c = 100$              & \textbf{\boldmath $0.05209$} \\ \hline
		Clustered ER network & $N_c = 100$              & \textbf{\boldmath $0.04964$} \\ \hline
		Clustered SF network & $N_c = 100$              & \textbf{\boldmath $0.05203$} \\ \hline
	\end{tabular}
\end{table}
\subsection{mpact of Network Topology on Reservoir State Dynamics and Correlations}
\begin{figure*}[!htbp] 
	\centering
	\includegraphics[width=1.0\linewidth]{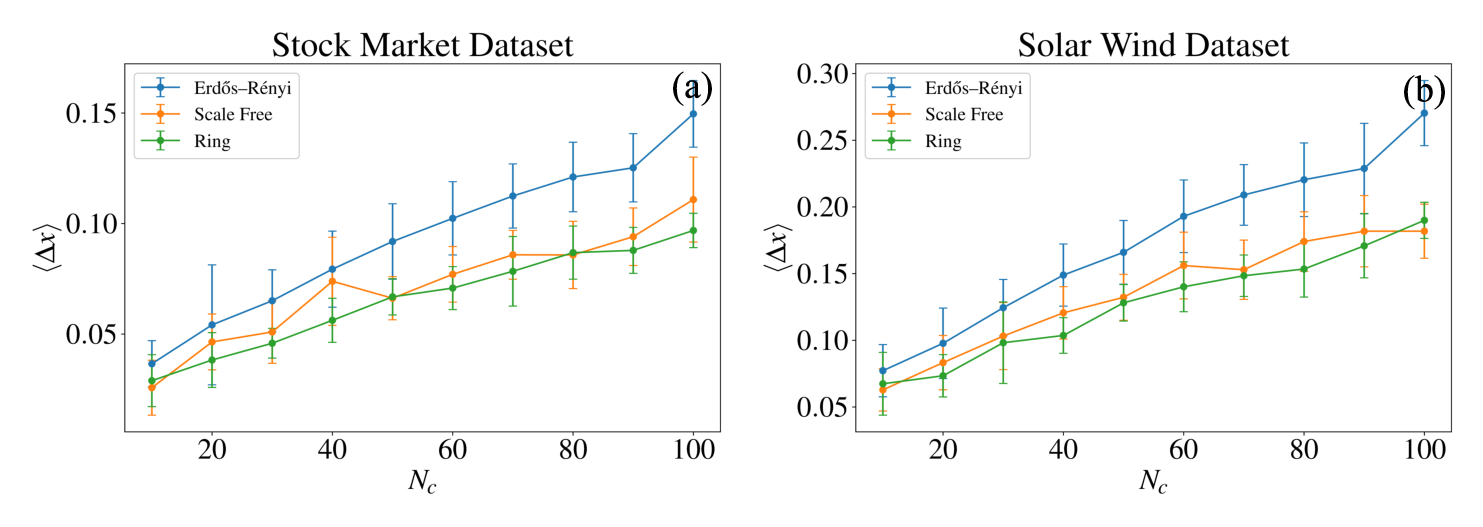}
	\caption{Average state change ($\langle \Delta x \rangle$) as a function of the number of neurons per cluster ($N_c \in [10, 100]$) for two complex datasets: (a) stock market and (b) solar wind. Each data point represents the mean value computed over 50 independent realizations, providing insight into the degree of state redundancy across different network topologies.}%
	\label{state_change}
\end{figure*}
We have already demonstrated how the proposed algorithm performs on two complex datasets, the stock market and solar wind time series. Next, we investigate how the underlying network topology influences the reservoir’s state dynamics. As discussed in Section~\ref{sec41}, when $N_c > 40$ and the inter-cluster connectivity becomes weak, the overall prediction performance tends to deteriorate. Because the ring topology exhibits a highly structured connectivity pattern and the SF topology features hub-dominated links, it is important to assess the degree of redundancy in information transmission as $N_c$ increases. To this end, we quantify the state redundancy by computing the average state change ($\langle \Delta x \rangle$), which measures the mean magnitude of state updates between consecutive time steps, expressed as:
\begin{equation}
	\label{eq:avg_state_change}
	\langle \Delta x \rangle = \frac{1}{T-1} \sum_{t=1}^{T-1} 
		\left\| \mathbf{x}(t+1) - \mathbf{x}(t) \right\|,
\end{equation}
where $\mathbf{x}(t) \in \mathbb{R}^{N}$ denotes the reservoir state vector at time step $t$, $N$ is the total number of reservoir neurons, and $\| \cdot \|$ represents the Euclidean norm.  This metric quantifies the dynamical responsiveness of the network: a larger value of $\langle \Delta x \rangle$ indicates richer and more diverse state transitions, reflecting stronger nonlinearity and reduced redundancy, whereas a smaller value corresponds to synchronized neuron evolution and greater redundancy in state updates~\cite{jaeger2001echo}.

\begin{figure*}[!htbp] 
	\centering
	\includegraphics[width=1.0\linewidth]{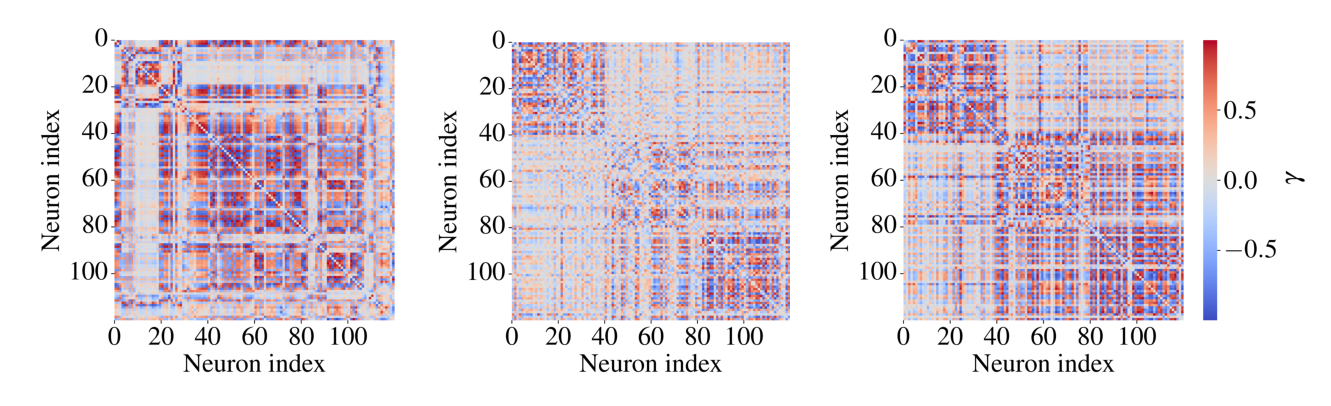}
	\caption{Correlation matrices of reservoir states for $N_c =$ 40 obtained from the solar wind dataset, illustrating (a) clustered ring, (b) clustered ER, and (c) clustered SF network topologies. The color bar, denoted by $\gamma$, represents the correlation coefficient, indicating the degree of similarity among reservoir node activations within and across clusters. To enhance visual contrast and better depict inter-neuronal correlations, the diagonal elements of the matrices are set to zero instead of one.}%
	\label{state_corr}
\end{figure*}

To evaluate this behavior, we compute $\langle \Delta x \rangle$ for different cluster sizes $N_c \in [10,100]$ using both the stock market [Fig.~\ref{state_change}(a)] and solar wind [Fig.~\ref{state_change}(b)] datasets. As shown, $\langle \Delta x \rangle$ increases with $N_c$, with the ER reservoir consistently producing the highest values (depicted in blue), signifying enhanced dynamical diversity and minimal redundancy among neurons. In contrast, the ring topology (depicted in green) exhibits smaller $\langle \Delta x \rangle$ values as $N_c$ increases, owing to its regular, linear structure, which slows information propagation and results in greater state redundancy. For the SF network, hub nodes promote local synchronization, resulting in partially correlated and redundant subgroups of neurons. However, despite this higher redundancy, the SF topology remains advantageous for tasks involving long-term dependencies, where the persistence of correlated states can enhance predictive accuracy.

To further substantiate these observations, we compute correlation matrices of the reservoir states for the solar wind dataset with $N_c = 40$, as illustrated in Fig.~\ref{state_corr}. In the ring topology [Fig.~\ref{state_corr}(a)], strong correlations among numerous neurons indicate high redundancy in the reservoir dynamics. For the clustered ER network [Fig.~\ref{state_corr}(b)], localized regions of high correlation appear within certain cluster subgroups, while inter-cluster correlations remain weak, implying diverse state evolution and lower redundancy. The SF topology [Fig.~\ref{state_corr}(c)] exhibits a heterogeneous correlation pattern characterized by strong intra-cluster correlations around hub nodes and weaker inter-cluster correlations elsewhere, consistent with its scale-free architecture. Together, these analyses reinforce that the choice of network topology critically influences the richness, independence, and redundancy of reservoir dynamics, thereby shaping the CESN’s capacity to capture complex multivariate temporal dependencies.
\section{Chaotic system}
\label{sec5}

\begin{figure}[!htbp] 
	\centering
	\includegraphics[width=0.6\linewidth]{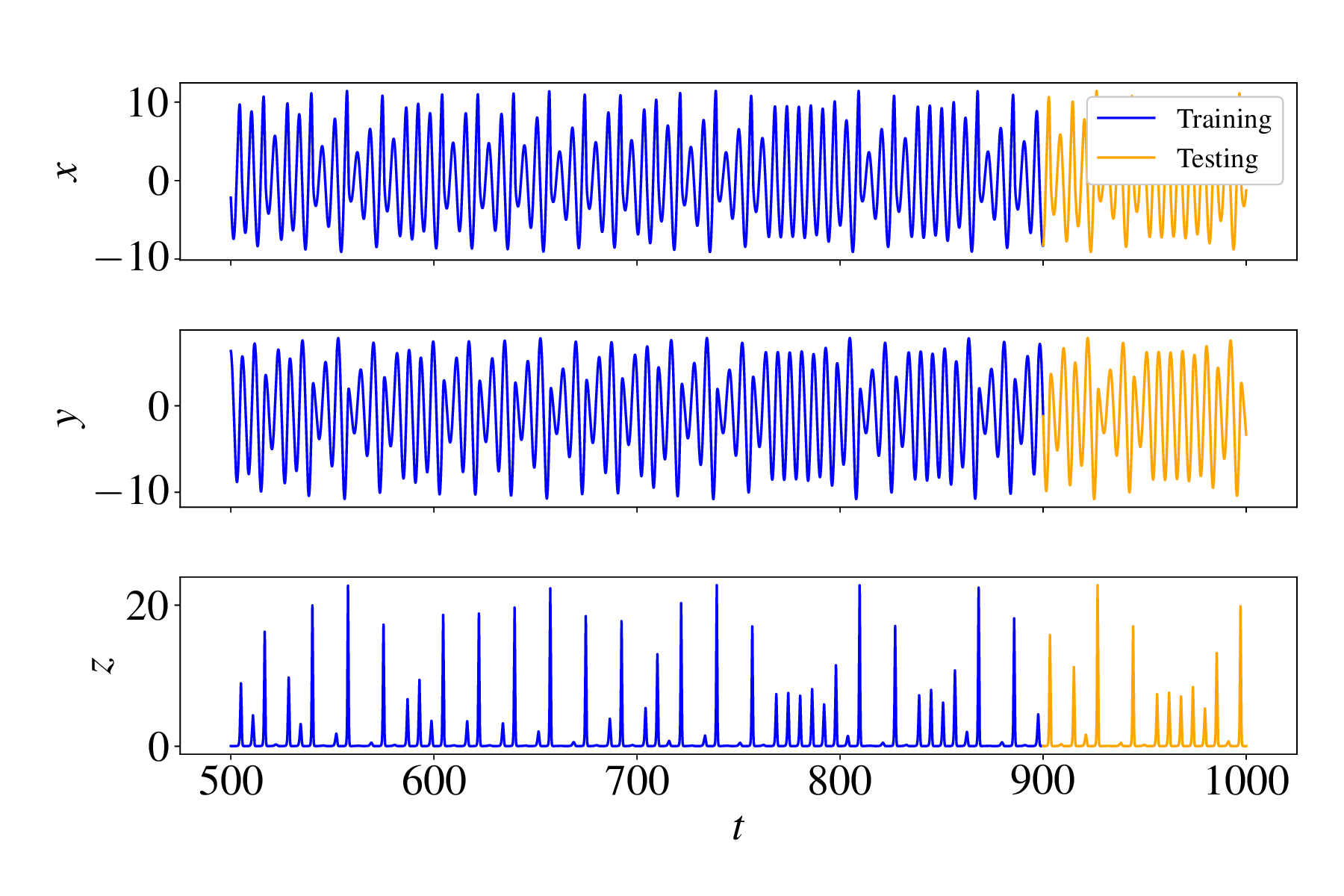}
	\caption{Time series data generated from the numerical simulation of the R\ "ossler system. The dataset is partitioned into an 80\% training set (shown in blue) and a 20\% test set (shown in orange) for evaluating the model's predictive performance.}
	\label{ross} 
\end{figure}

\begin{figure}[!htbp] 
	\centering
	\includegraphics[width=0.5\linewidth]{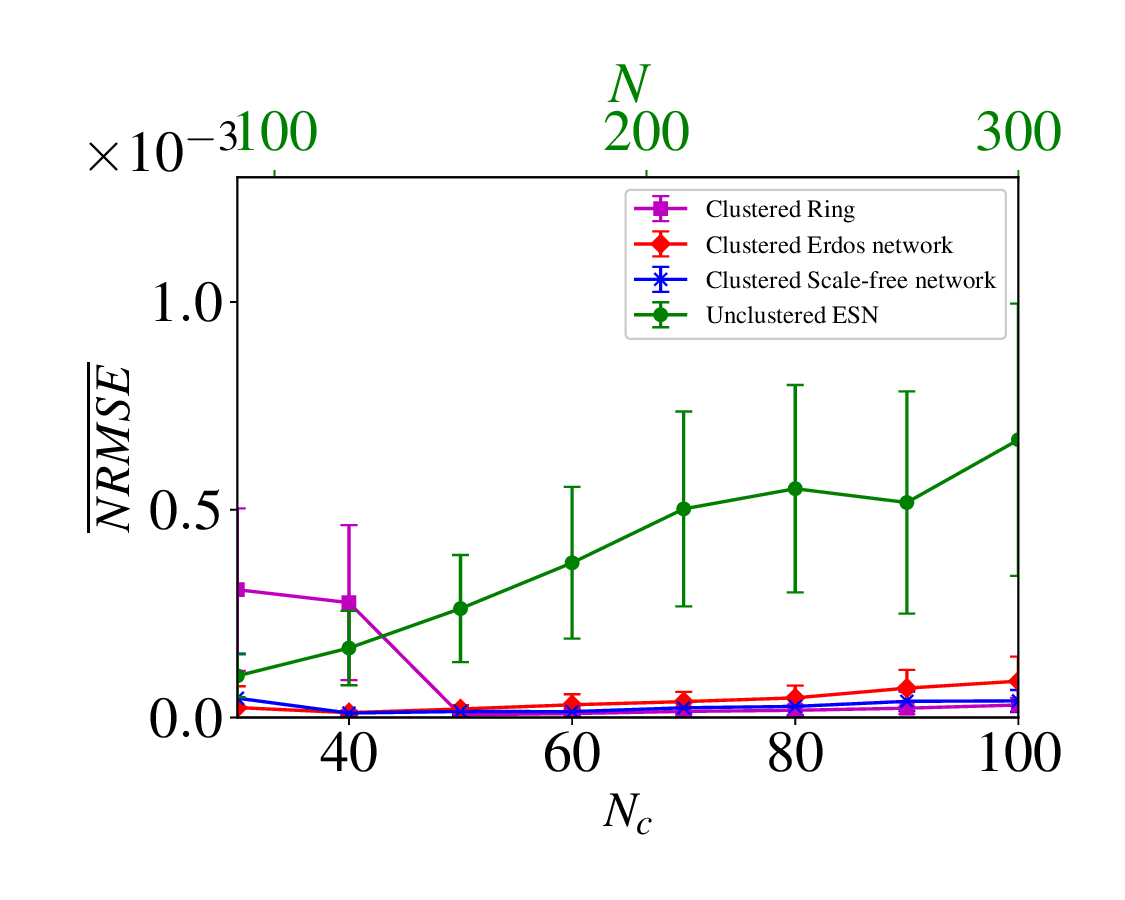}
	\caption{Average NRMSE ($\overline{\text{NRMSE}}$) plots for the R\"ossler time series, illustrating the effect of varying $N$ in unclustered ESNs (indicated by green $\times$-markers) and $N_c$ in CESNs. The spectral radius is fixed at $\rho = 0.9$ for all configurations.}
	\label{rossler}%
\end{figure}
To further evaluate the robustness and generalizability of the proposed algorithm, we apply it to a well-known chaotic system, the R\"ossler system. This system exhibits three dynamically evolving state variables, which naturally motivates the use of $m$ = 3 clusters, each corresponding to one of the system's variables.

The governing equations of the R\"ossler system are defined as:
\begin{equation}
	\begin{aligned}
		\frac{dx}{dt} &= -y - z, \\
		\frac{dy}{dt} &= x + a y, \\
		\frac{dz}{dt} &= b + z(x - c),
	\end{aligned}
\end{equation}
where $x$, $y$, and $z$ are the system's state variables, and the parameters are fixed at $a$ = 0.2, $b$ = 0.2, and $c$ = 5.7. The equations are numerically integrated using the fourth-order Runge--Kutta method with a time step of $ h = 0.01$. The initial conditions are set to $x$ = 1.0, $y$ = 1.0, and $z$ = 1.0. After discarding transients, the remaining time series is divided into training and test sets as illustrated in Fig.\ref{ross}. Prior to prediction, the data is normalized to fit the reservoir's operating range. The $\overline{\text{NRMSE}}$ values for different network configurations are presented in Table \ref {nrmse_ross}, with clustered ER and SF networks exhibiting superior predictive performance.

\begin{table}[!htbp]
	\centering
	\renewcommand{\arraystretch}{1.2} 
	\caption{Average NRMSE ($\overline{\text{NRMSE}}$) errors for R\"ossler multivariate time series prediction using an 80\%-20\% train-test split.}
	\label{nrmse_ross}
	\begin{tabular}{|l|c|c|}
		\hline
		\textbf{Method} & Reservoir size & \textbf{$\overline{\text{NRMSE}}$} \\ \hline \hline
		Conventional ESNs \cite{jaeger2002adaptive}    & $N = 300$               & \textbf{\boldmath $3.4 \times 10^{-3}$}\\ \hline
		AEESNs  \cite{xu2016adaptive}              & $N = 100$               & $2.6 \times 10^{-3}$ \\ \hline
		DeepESNs \cite{malik2016multilayered}             & $N = [100,100]$  & $2.3 \times  10^{-4}$ \\ \hline
		BrESNs\cite{yao2019broad}                & $N = [100,100,100]$ & $3.12  \times 10^{-3}$ \\ \hline
		Clustered Ring ESNs   & $N_c = 100$              & \textbf{\boldmath $6.3 \times 10 ^{-4}$} \\ \hline
		Clustered ER network & $N_c = 100$              & \textbf{\boldmath $8.89 \times 10 ^{-5}$} \\ \hline
		Clustered SF network & $N_c = 100$              & \textbf{\boldmath $2.4 \times 10 ^{-5}$} \\ \hline
	\end{tabular}
\end{table}

We then explore the effect of varying the $N$ and $N_c$, as depicted in Fig.~\ref{rossler}. Across multiple independent realizations, we observe that the clustered architectures maintain consistent performance as $N$ increases. For small cluster sizes ($N_c < 40$), the clustered ring topology underperforms relative to the conventional ESN. However, beyond this threshold, the prediction error stabilizes. In contrast, both the clustered ER and SF networks consistently outperform all other configurations across the entire range of $N_c$ values.

\begin{figure*}[!htbp] 
	\centering
	\includegraphics[width=1.0\linewidth]{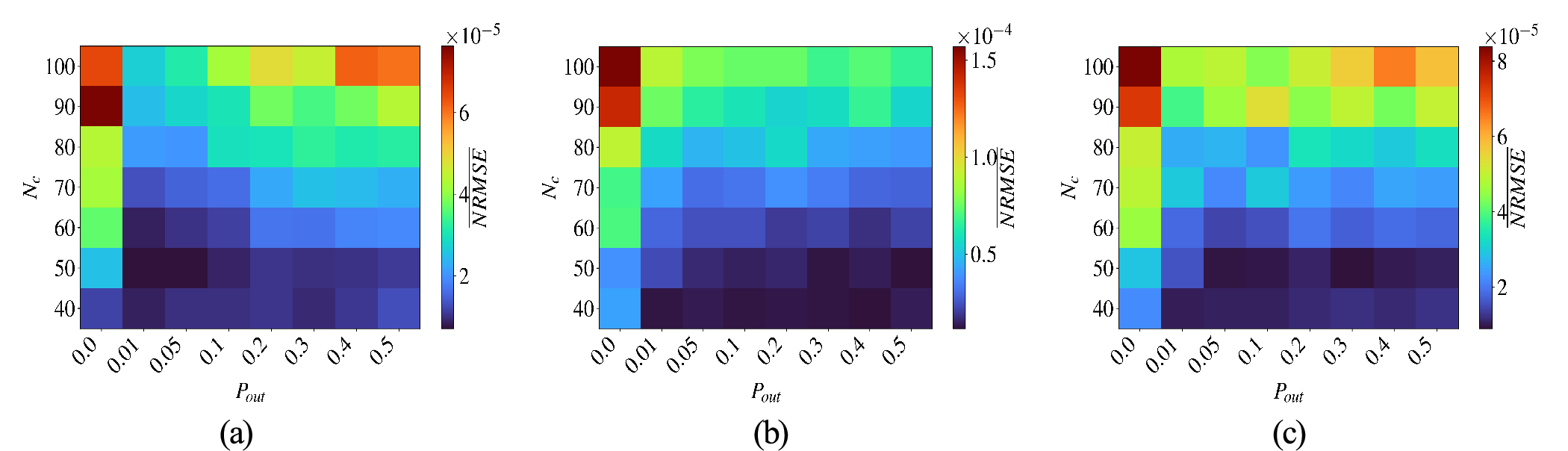}
	\caption{Two-parameter analysis of $P_{\text{out}}$ versus $N_c$ for the R\"ossler time series dataset. The color bar represents $\overline{\text{NRMSE}}$, highlighting both optimal prediction regions and the limitations of increasing inter-cluster connectivity. All other parameters are fixed as listed in Table~\ref{table}. Each parameter combination is evaluated over 50 independent realizations. Results are shown for the three clustered topologies: (a) ring, (b) ER, and (c) SF networks. }%
	\label{rossler_bar}
\end{figure*}

To further highlight the role of inter-cluster connectivity, we analyze the performance landscape of different topologies with respect to $N_c$ and $P_{\text{out}}$, the probability of connections between clusters. As shown in Fig. \ref{rossler_bar}, weak inter-cluster coupling substantially improves performance across all topologies. Specifically, in the ring topology (Fig.\ref{rossler_bar}(a)), weak interconnections ($P_{\text{out}} = 0.01$ to $0.1$) lead to better generalization compared to both isolated and highly interconnected clusters. Similar trends are observed in the ER and SF networks (Fig.~\ref{rossler_bar}(b–c)), where mild inter-cluster connectivity significantly enhances prediction, especially for smaller cluster sizes.

Overall, our findings confirm that CESNs, particularly those employing ER and SF topologies, offer significant advantages over conventional ESNs for learning complex temporal dynamics. Their ability to maintain stable and accurate predictions across a wide range of parameters makes them well-suited for real-world applications involving noisy, nonlinear, and multivariate time series, as well as chaotic systems like the R\"ossler attractor.

\section{Discussion and conclusion}
\label{sec6}
The introduction of RNN within the reservoir framework significantly simplified training by requiring optimization only at the output layer. This reduction in trainable parameters resulted in a significant decrease in computational cost. ESNs, by maintaining the echo state property, demonstrated superior predictive capabilities and have been successfully applied to various temporal and classification tasks. Inspired by the brain's modular structure, we propose a clustered reservoir architecture where nodes are grouped into distinct clusters. This approach enhances the prediction accuracy of multivariate time series, which poses a significant challenge for conventional random ESNs. In the proposed method, each input variable from the multivariate time series is mapped to a specific cluster within the reservoir, and training is performed on the corresponding cluster in the readout layer. This cluster-based methodology improves the model's capacity to capture complex temporal dependencies compared to unstructured random ESNs.

Given that reservoir topology directly affects predictive performance, this work also investigates the influence of different network architectures by first conducting a comprehensive analysis of inter-cluster connectivity. This preliminary investigation, followed by an evaluation of model accuracy and non-predictive indicators such as training time and memory utilization, enables a systematic identification of the key characteristics that define each network topology. To evaluate the effectiveness of the proposed approach, we applied it to two real-world multivariate time-series datasets, solar wind and stock market data, both of which are inherently noisy and characterized by random fluctuations. Furthermore, the influence of reservoir topology on state redundancy is examined through the average state-change metric, providing insights into how correlated reservoir states can either hinder or enhance predictive performance. In addition, we tested the algorithm on a chaotic benchmark system, specifically the R\"ossler attractor. Across all these diverse and challenging tasks, the proposed CESNs consistently outperformed conventional ESNs, even when utilizing a minimal number of reservoir nodes. The clustered architectures exhibited stable predictive performance across a wide range of spectral radii and cluster sizes. Notably, even the simplest clustered topology, the ring network, surpassed conventional ESNs in prediction accuracy. Among the different configurations explored, the clustered ER and SF networks consistently delivered the most accurate results across the multivariate datasets.

The importance of inter-cluster connections was thoroughly investigated using a two-parameter analysis involving  $N_c$ and $P_{\text{out}}$. This analysis revealed key trade-offs in network design: in certain scenarios, employing fewer nodes per cluster with weaker interconnections led to improved prediction accuracy. Conversely, when clusters contain a larger number of nodes, the reservoir learns more effectively in isolation, and excessive interconnections may degrade performance. However, this behaviour is not universal and varies depending on the nature of the dataset. For instance, in the case of the chaotic R\"ossler system, increasing the number of nodes per cluster and encouraging inter-cluster communication actually improved predictive accuracy, underscoring the dataset-specific nature of optimal connectivity.

The demonstrated superiority of CESNs across diverse domains highlights their potential as a powerful framework for time series modeling. Future research could explore the application of clustered reservoirs to other complex and high-dimensional time series prediction problems. Adding to it, some of the research limitations and possible future directions are listed below:

\subsection{Research Limitations \& Possible Future Directions}
\begin{itemize}
	\item The proposed CESN framework demonstrates strong predictive capability with a relatively small number of neurons per cluster ($N_c$) across different topologies, particularly when inter-cluster connections are weak. However, as $N_c$ increases, the performance of independent clusters begins to surpass that of interdependent ones, and the overall results become increasingly sensitive to the underlying network topology. Therefore, future research should focus on enhancing the model’s robustness by developing a topology-invariant framework that maintains consistent performance across different architectures. Such an approach could enable a more universal and adaptive clustered network design optimized for predicting diverse multivariate time series.
	
	\item Although the CESN methodology has proven effective for time-series prediction, the optimal level of inter-cluster connectivity cannot be universally fixed, as it is inherently dependent on the characteristics of the dataset under study. Consequently, future investigations should explore data-driven, adaptive mechanisms for dynamically regulating inter-cluster connectivity based on the temporal and statistical properties of the input signals, thereby enhancing model generalization and self-organization.
\end{itemize}

\section*{Acknowledgments}
The research contributions of S.H., R.S., and V.K.C. are part of a project funded by the SERB-CRG (Grant No. CRG/2022/004784). The authors gratefully acknowledge the Department of Science and Technology (DST), New Delhi, for providing computational facilities through the DST-FIST program under project number SR/FST/PS-1/2020/135, awarded to the Department of Physics.\\

\noindent {\bf Author Contributions} All the authors contributed equally to the preparation of this manuscript.\\

\noindent {\bf Data Availability Statement} The authors confirm that the data supporting the findings of this study are available within the article.\\

\noindent {\bf Conflict of interest} The authors declare that they have no conflict of interest.

\bibliographystyle{unsrt}
\bibliography{cas-refs}

\end{document}